%% file: NA_v1.tex
\documentclass{nature}
\input{preamble.tex}

\usepackage{lscape}
\usepackage{mathtools}
\usepackage{gensymb}
\usepackage{threeparttable}

\title{Further evidence for a population of dark-matter-deficient dwarf galaxies}

\author{ Qi Guo$^{\rm * 1,2,3}$; Huijie Hu$^{\rm 1,2}$; Zheng Zheng$^{\rm 1,4}$; Shihong Liao$^{1,3}$; Wei Du$^{\rm 1,5}$; Shude Mao$^{\rm 6,1}$; Linhua Jiang$^{\rm 7}$; Jing Wang$^{\rm 7}$; Yingjie Peng$^{\rm 7}$; Liang Gao$^{\rm 1,2,3}$; Jie Wang$^{\rm 1,2,3}$; Hong Wu$^{\rm 1,2,5}$}

\begin{document}
\maketitle

\begin{affiliations}
\item National Astronomical Observatories, Chinese Academy of Science, 20A Datun Road, Chaoyang District, Beijing, 100101, China
\item University of Chinese Academy of Sciences, No. 19 A Yuquan Road, Beijing, 100049, China
\item Key Laboratory for Computational Astrophysics, National Astronomical Observatories, Chinese Academy of Sciences, Beijing 100101, China
\item Key Laboratory of
The Five-hundred-meter Aperture Spherical radio Telescope, National Astronomical Observatories, Chinese Academy of Sciences
\item Key Laboratory of Optical
Astronomy, National Astronomical Observatories, Beijing, 100101, China
\item Physics Department and Tsinghua Center for Astrophysics, Tsinghua University, Beijing 100084, China
\item Kavli Institute for Astronomy and Astrophysics, Peking University, Beijing 100871, China
\end{affiliations}

\begin{abstract}
In the standard cosmological model, dark matter drives the structure formation and constructs potential wells within which galaxies may form. The baryon fraction in dark halos can reach the universal value (15.7\%) in massive clusters and decreases rapidly as the mass of the system decreases\cite{White93,Fukugita98}. The formation of dwarf galaxies is sensitive both to baryonic processes and the properties of dark matter owing to the shallow potential wells in which they form. 
In dwarf galaxies in the Local Group, dark matter dominates the mass content even within their optical-light half-radii (r$_{\rm e}$ $\rm \sim 1$ kpc)\cite{Walker12,Strigari08}.
However, recently it has been argued that not all dwarf galaxies are dominated by dark matter\cite{van18,van19,Oman16}. Here we report 19 dwarf galaxies that could consist mainly of baryons up to radii well beyond r$_{\rm e}$, at which point they are expected to be dominated by dark matter. Of these, 14 are isolated dwarf galaxies, free from the influence of nearby bright galaxies and high dense environments. This result provides observational evidence that could challenge the formation theory of low-mass galaxies within the framework of standard cosmology. Further observations, in particular deep imaging and spatially-resolved kinematics, are needed to constrain the baryon fraction better in such galaxies.
\end{abstract}

In the standard cosmology\cite{Planck18}, dark matter contributes around 26.6\% of the Universe, while baryons only occupy 4.9\%. The baryonic fraction is about the universal value, 15.7\%, in galaxy clusters, and decreases rapidly towards lower masses. In low mass systems, the potential wells are shallow so that various baryonic processes, including background UV ionization and stellar feedback, could either prevent gas infall or expel gas out of the systems, leading to a low baryonic fraction. Observationally, it infers a typical mass-to-light ratio of 10--1000, using stellar dynamics of the dwarf satellite galaxies in the Local Group\cite{Simon07,Martin07}. Such values are supported by modern simulations\cite{Sawala15}. However, recently two dwarf galaxies were found to have baryonic mass comparable to the dark matter mass up to several kiloparsecs using the dynamics of their HI gas\cite{Oh15,Oman16}.  Similar results were reported for two ultra-diffuse dwarf galaxies in the NGC1052 group \cite{van18, van19} using the dynamics of their globular clusters, of which the NGC1052-DF4 dwarfs may have almost no dark matter at all.  Such relation between dark matter and baryons is also revealed in the population of tidal dwarf galaxies\cite{Bournaud07,Fritz18}.  In all cases, these galaxies reside in (relatively) high density regions, and the environmental stripping/interaction might play a role. 

Here we search for baryon dominated dwarf galaxies (r-band absolute magnitude, M$_{\rm r}>-18$) in the field, using the catalogue of Arecibo Legacy Fast ALFA (ALFALFA) $\alpha$.40\cite{Haynes11}, which matches the HI selected galaxies to the Seventh Data Release of Sloan Digital Sky Survey (SDSS DR7)\cite{Abazajian09} PhotoPrimary optical catalogue. Signal-to-noise cut at HI spectrum, SNR$_{\rm HI}>10$, is imposed. We further remove sources with suspicious optical images and HI spectra, and sources which have multiple Galaxy Evolution Explorer(GALEX)\cite{Martin05} images. The photometric and geometric properties are re-calculated for low surface brightness galaxies for SDSS DR7 pipeline systematically over-estimate the sky background for such systems\cite{Lisker2007,Zheng2015}. In the lack of HI image, we use the r-band b/a ratio to correct for the inclination. In order to minimize the mis-alignment effect between the HI content and the r-band image, we select galaxies with r-band b/a ratio between 0.3 and 0.6, leading to a final parent sample of 324 galaxies.  
 The stellar mass of each galaxy is derived from the optical photometry from SDSS and the distance measured by the ALFALFA\cite{Bell03}. Dynamical mass is estimated using the circular velocity V$_{\rm HI}$ which is derived from 20\% of the HI line width (w$_{20}$), and the HI radius (r$_{\rm HI}$, defined at HI surface density =  1M$_{\odot}$pc$^{-2}$) which is converted from the HI mass based on the tight relation between HI radius and HI mass\cite{Wang16}. We verify the accuracy of such defined r$_{\rm HI}$ and V$_{\rm HI}$ using dwarf galaxies in the Local Irregulars That Trace Luminosity Extremes The HI Nearby Galaxy Survey (LITTLE THINGS), where the HI images have been well resolved and HI dynamics been detailed analyzed. Details are given in the Method. 
 
Figure 1 shows images and spectra of two example dwarf galaxies: AGC 213086 and AGC 8915. The r-band luminosity of AGC 213086 is $3.07\times10^{42}$ erg s$^{-1}$, corresponding to a stellar mass of $5.51_{-2.32}^{+4.02} \times 10^{8}$M$_{\odot}$. The HI mass is $2.45_{-0.09}^{+0.11} \times 10^{9} $M$_{\odot}$, corresponding to a radius of r$_{\rm HI} = 14.37 \pm 1.023$ kpc. This radius is much larger than the half light radius in the optical wavelength, r$_{\rm e} \sim 3.39$ kpc, extending to the dark matter dominated region if AGC 213086 were a typical dwarf galaxy. The total baryonic mass (M$_{\rm bary}$) within r$_{\rm HI}$ is $3.82_{-0.32}^{+0.34} \times 10^{9} $M$_\odot$. The circular velocity is $43.47 \pm 2.40$ km/s and the dynamical mass (M$_{\rm dyn}$) enclosed within the r$_{\rm HI}$ is $6.31_{-0.77}^{+0.89} \times 10^9$ M$_{\odot}$. Error bars account for various uncertainties in the data and in the method for the mass estimation. Assuming a typical NFW profile\cite{NFW97} the virial mass (M$_{\rm halo}$, the total mass within the virial radius, R$_{\rm vir}$, within which the mean density is 200 times the critical density) is $1.41 \times 10^{10}$M$_\odot$, and is $1.01 \times 10^{10}$M$_\odot$ if assuming a Burkert profile\cite{Burkert95}. The ratio between M$_{\rm dyn}$($<$r$_{\rm HI}$) and M$_{\rm bary}$($<$r$_{\rm HI}$) is less than 2, suggesting baryon dominates over dark matter within r$_{\rm HI}$. The baryonic fraction within R$_{\rm vir}$ is 0.271 (and 0.376 if assuming a Burkert profile), well above the universal value. 
In comparison, we show at the bottom a `typical' dwarf galaxies, AGC 8915. It has M$_{\rm bary}$($<$r$_{\rm HI}$) = M$_{\rm gas}$($<$r$_{\rm HI}$) + M$_{\rm star}$ ($<$r$_{\rm HI}$)= $3.14_{-0.17}^{+0.18} \times 10^9$M$_{\odot} \ll$ M$_{\rm dyn}$($<$r$_{\rm HI}$)= $2.647_{-0.18}^{+0.20}  \times 10^{10}$ M$_{\odot}$ and M$_{\rm halo}=$ $1.83\times 10^{11}$ M$_{\odot}$ (NFW profile ) / $1.04\times 10^{11}$ M$_{\odot}$ (Burkert profile). Its M$_{\rm dyn}$-to-M$_{\rm bary}$ ratio within r$_{\rm HI}$ is $\sim$10, and the baryonic fraction within R$_{\rm vir}$ is 0.0171 (NFW profile) / 0.0301 (Burkert profile), well below the universal value. 

The main statistical results are presented in Figure 2. We find 19 dwarf galaxies out of 324 parent dwarf galaxies, a non-negligible fraction, with M$_{\rm dyn}$($<$r$_{\rm HI}$) $<$ 2 $\times$ M$_{\rm bary}$($<$r$_{\rm HI}$), i.e. M$_{\rm DM}$($<$r$_{\rm HI}$) $< $M$_{\rm bary}$($<$r$_{\rm HI}$). These 19 dwarf galaxies have more extended r$_{\rm e}$ distribution and  r$_{\rm HI}$ distribution compared to the parent sample. For each of them, r$_{\rm HI}$ is much larger than r$_{\rm e}$, suggesting its dynamics is estimated at rather large radius where dark matter is expected to be dominant.

The median M$_{\rm dyn}$($<$r$_{\rm HI}$)/M$_{\rm bary}$($<$r$_{\rm HI}$) (here after M$_{\rm dyn}$/M$_{\rm bary}$)  of the parent sample is 7.36 with the one-$\sigma$ dispersion of [3.85, 11.03], well below the the typical values found in the dwarf galaxies in the Local Group (right panel of Figure 2). There is a clear excess of the baryonic dominated dwarf galaxies (here after BDDGs) compared to a Gaussian distribution in the log M$_{\rm dyn}$/M$_{\rm bary}$, suggesting BDDGs to be a distinct galaxy population rather than statistical outliers. 

 Figure 3 shows that as the parent sample, for most BDDGs M$_{\rm dyn}$($<$r$_{\rm HI}$) is dominated by gas masses, so that the potential mis-matching between the optical and HI sources could not affect the results substantially. The parent sample generally follows the same Tully-Fisher relation as spiral galaxies, yet BDDGs tends to have much lower circular velocities for given luminosities. It is the low circular velocities rather than systemic uncertainties (e.g. geometric structures, distance estimates and etc.) that leads to their small M$_{\rm dyn}$/M$_{\rm bary}$ values.   
 Assuming the NFW profile, we find the BDDGs reside in halos as small as $\rm 10^9 \sim 10^{11}$ M$_{\odot}$ (Figure 4), consistent with the  mass range if assuming the Burkert profile. Their baryon fractions within R$_{\rm vir}$ are mostly well above the universal value. This is unexpected for these systems have rather shallow potential well and are thus harder to maintain their baryons against various feedback.
Hydro-dynamical simulations of galaxy formation show that interactions of galaxies at high density region could be responsible for the formation of baryonic dominated tidal dwarf galaxies, and the environmental stripping could play a role in forming the BDDGs around clusters\cite{Jing19,Ogiya18}. However, only 5 of the 19 BDDGs are within/close to groups/clusters. The rest of the 14 BDDGs reside at distances farther than 3 times the virial radii of any surrounding groups/clusters (right panel in Figure 4). This makes the environmental stripping and origin from interaction of galaxies very unlikely.  Our results suggest that a population of dwarf galaxies could form in a particular way such that much less dark matter is required than for those in the Local Group and those found in simulations.

One possible channel is that, under extreme conditions, strong supernova feedback could flatten the potential well by expelling gas. Along with this procedure dark matter could be removed to the outer region, in response to the gravity of the expelled gas. Further fueling cold gas might reach the potential well along filaments(cold flow), while dark matter grow in a more smooth and steady way. Under such special circumstance, it is possible to form BDDGs. Further studies with high resolution simulations are required to explore such possibility. A warm dark matter model could also likely lead to a relatively low density in the inner region of a given halo and thus could probably be more in line with the existence of the BDDGs. 

Alternatively, the systematic errors in inclination may lead to an under-estimate of the circular velocity and thus an under-estimate of the dynamical mass, for there could be mis-alignment between the optical inclination and the inclination of the HI kinematic field. Applying the same mis-alignment distribution between the optical images\cite{Hunter12} and the HI dynamics from the LITTLE THINGS and those from the most up-to-data  simulations\cite{Vogelsberger2014} to our sample, we find similar distributions of the M$_{\rm dyn}$/M$_{\rm bary}$, suggesting the existence of the BDDGs. In addition, most of our BDDGs are field dwarf galaxies for which less mis-alignment effect is expected. There could also exist systematic errors in optical inclinations, i.e. galaxies with high axis ratios could be  bars in face-on disks. Recent work \cite{Diaz-Garcia2016} shows a fraction of 10\% of such cases for dwarf galaxies, thought the fraction could be lower in low dense environments where most of our BDDGs are found. Detailed HI interferometric mapping of these galaxies in the future are required to rule out such possibilities.

\clearpage

\begin{figure*}
\centering
\includegraphics[width = 17.5 cm]{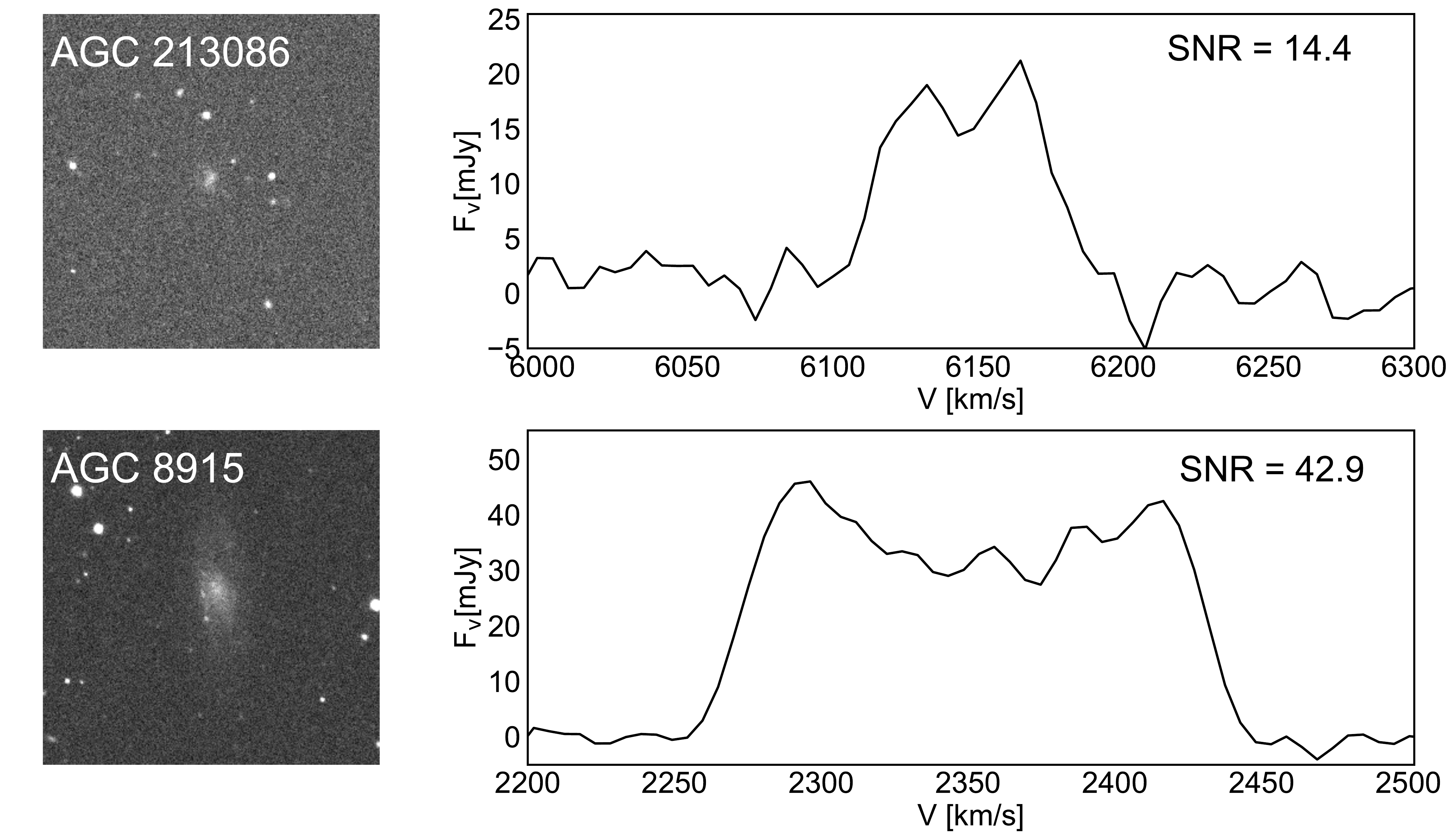} 
\caption{
{\bf Image and Spectra of two galaxy examples.} The square panels show the SDSS $\it r$-band image of two selected examples. Each panel spans 3'$\times$3'. The spectrum from ALFALFA of each object is shown next to the corresponding SDSS image, centered at the HI lines. The velocities are in unit of km/s. The spectral resolution of ALFALFA is 10 km/s. The signal-to-noise ratio is illustrated on the upper right corner of each spectrum. The upper one shows a dwarf galaxy with M$_{\rm bary}\sim $  M$_{\rm dyn}$, while the lower one shows a typical dwarf sample with M$_{\rm bary}\ll $  M$_{\rm dyn}$.}
\label{fig:Spectrua}
\end{figure*}

\begin{figure*}
\centering
\includegraphics[width=8cm]{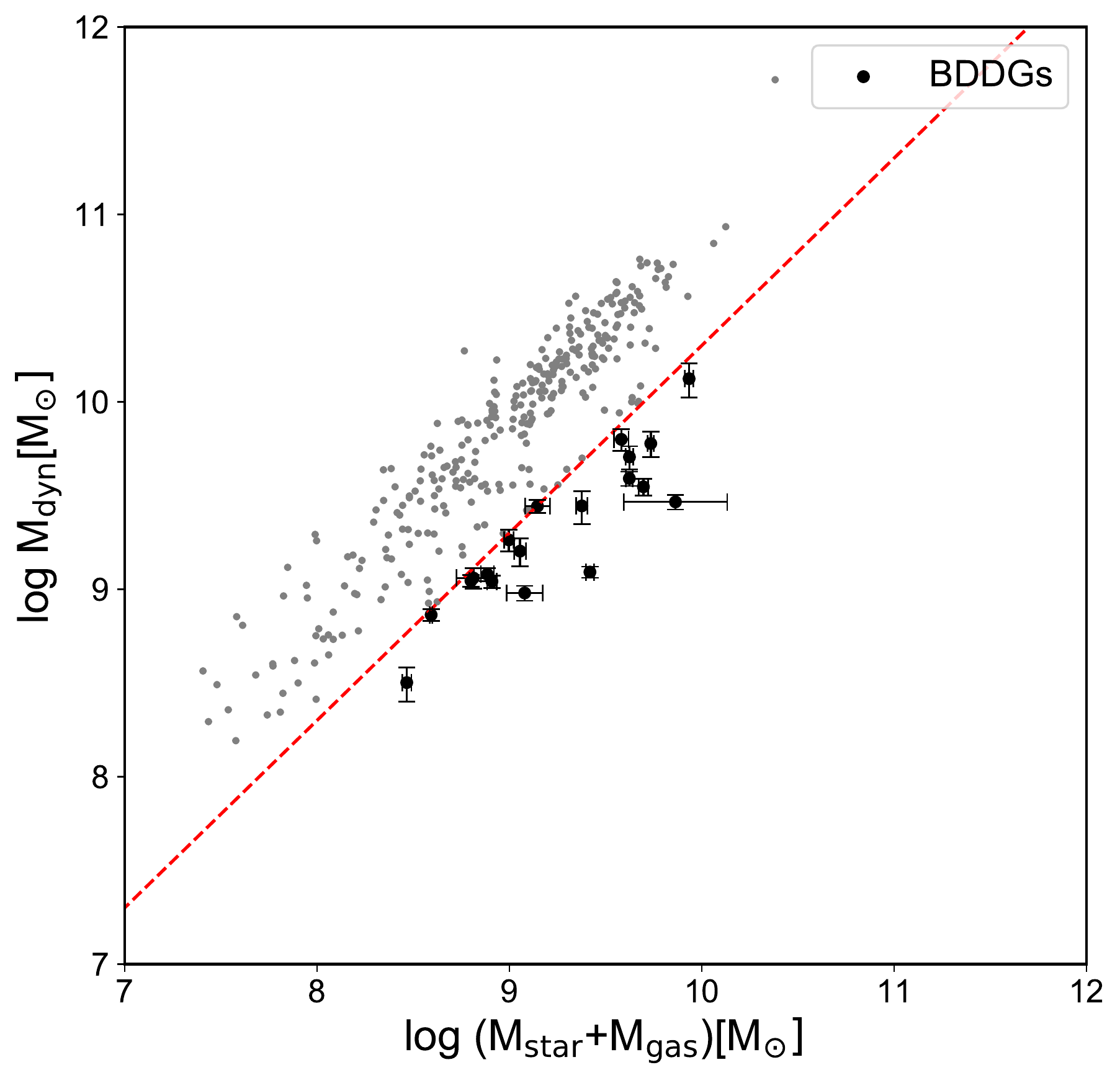}
\includegraphics[width=7.8cm]{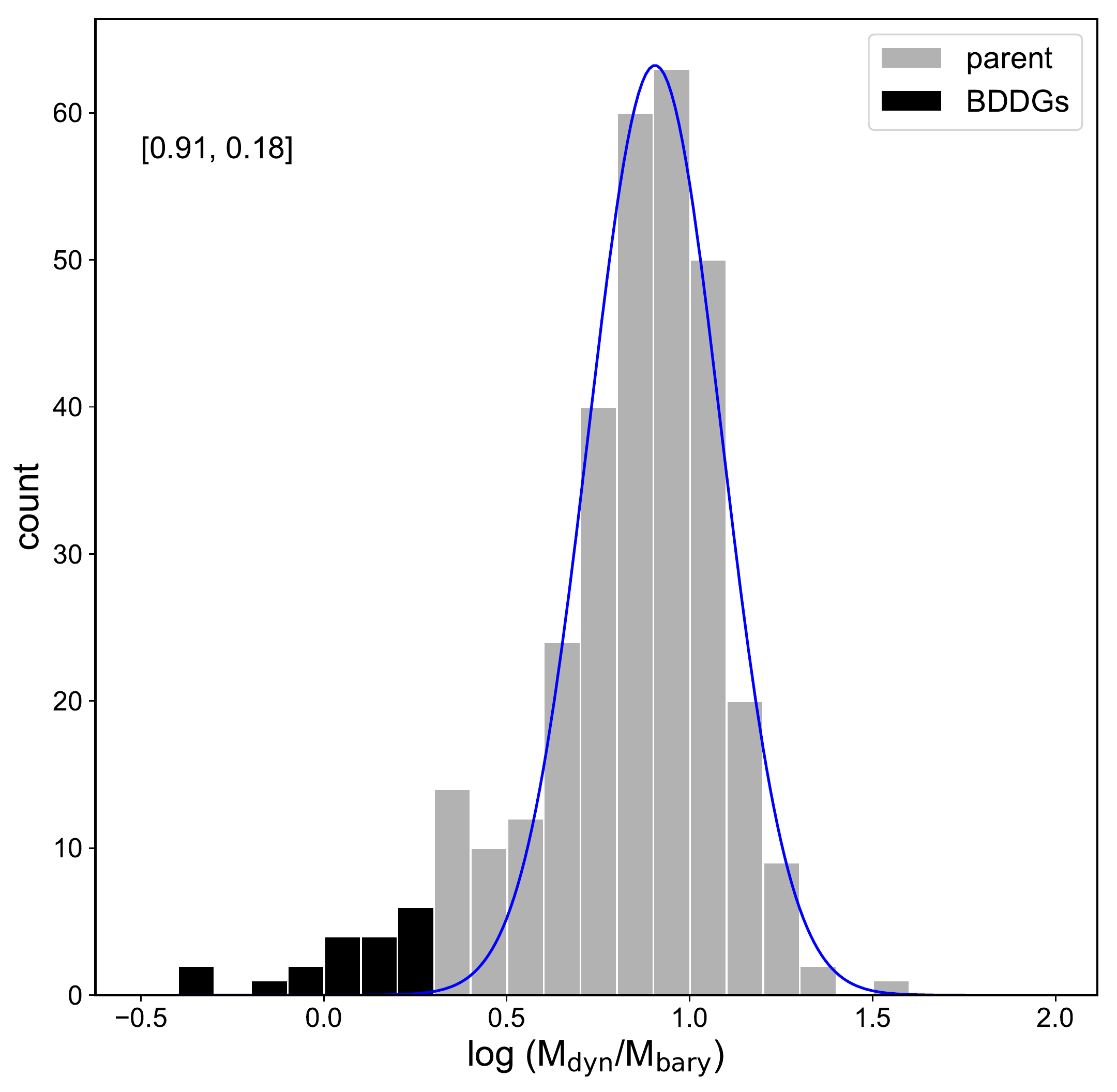}
\caption{
{\bf Comparison of baryonic mass and total mass enclosed within r$_{\rm HI}$.}
{\bf Left panel:} Baryonic mass vs. total mass enclosed within r$_{\rm HI}$. Grey dots are the parent sample with M$_{\rm r}>-18$, and black dots with error bars are the 19 baryonic dominated dwarf galaxies. Red dash line indicates M$_{\rm dyn}$ = $2 \times $M$_{\rm bary}$. Error bars account for various uncertainties in the data and in the method for the mass estimation.
{\bf Right panel:} distribution of log(M$_{\rm dyn}$/M$_{\rm bary}$) of baryonic dominated dwarf galaxies (black histogram) and the parent sample (grey histogram). Blue curve is the Gaussian fit to the parent sample, with the mean value of 0.91 and one $\sigma$ dispersion of 0.18. A clear excess of BDDGs is presented.
\label{fig:Dynamicalmass}
}
\end{figure*}

\begin{figure*}
\centering
\includegraphics[width=8cm]{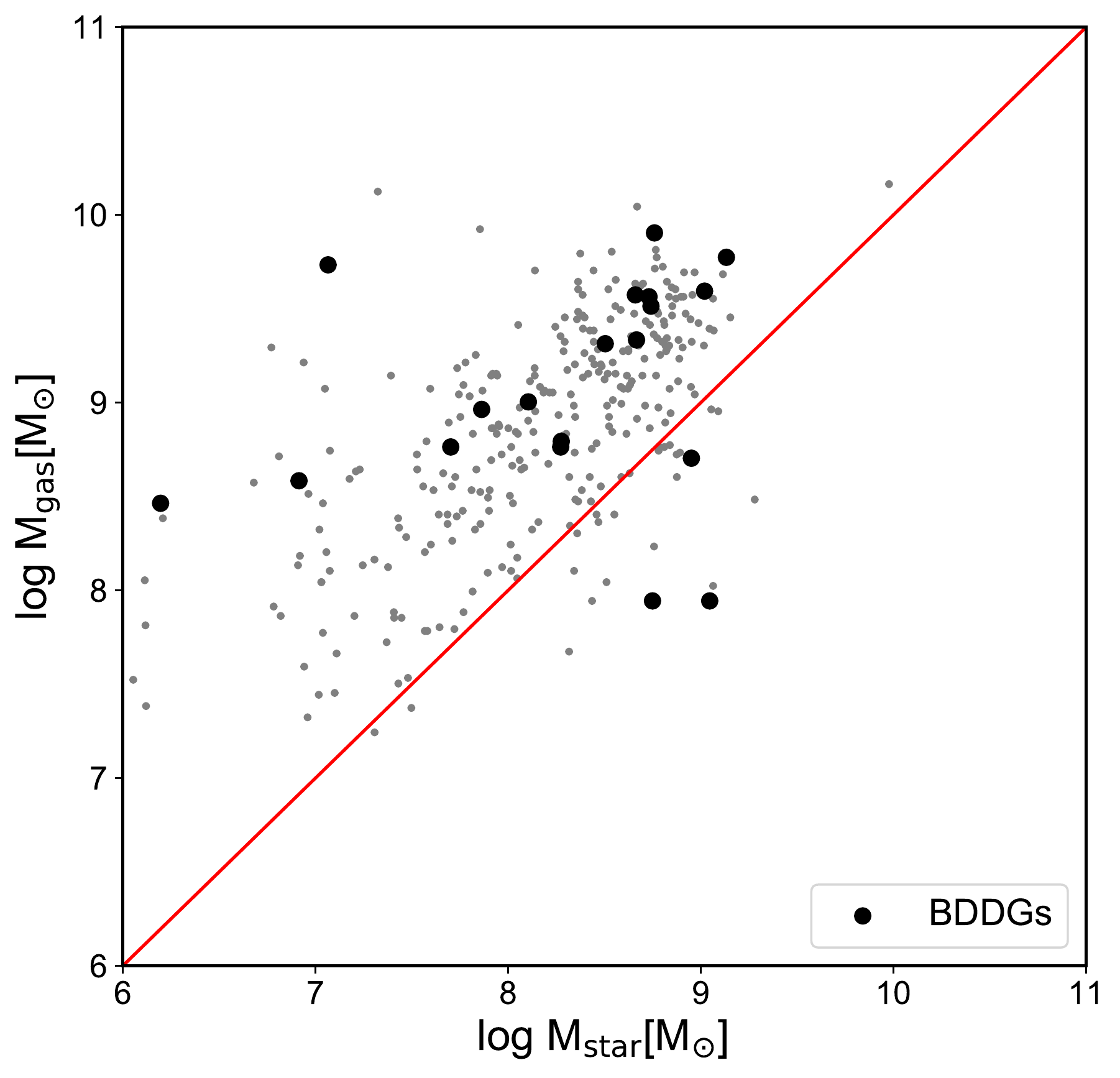}
\includegraphics[width=8cm]{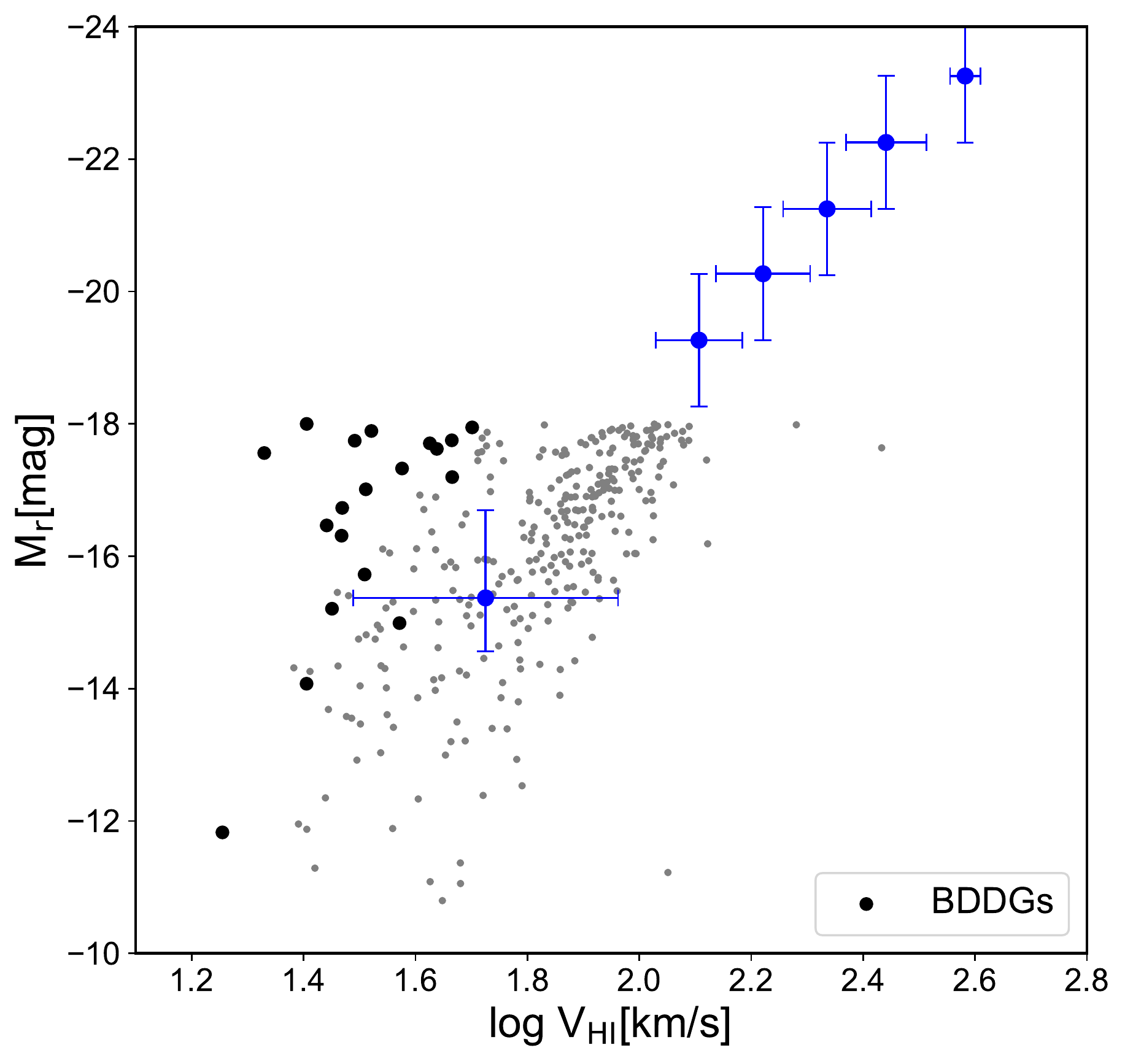}
\caption{{\bf Galaxy properties. Left panel}: Gas mass vs. stellar mass relation. Grey dots are for the parent sample, and black dots are for the baryonic dominated dwarf galaxies. Most of the parent galaxies have gas masses dominant over stellar masses. Their low M$_{\rm dyn}$/M$_{\rm bary}$ compared to those found in the dwarf galaxies in the Local Group can thus be partly explained by the removal of gas at high density region. {\bf Right panel}: r-band Tully-Fisher relation. V$_{\rm HI}$ is the rotation velocity measured using w$_{20}$ which also takes into account the inclination and the velocity dispersion. The Tully-Fisher relation for more massive galaxies are shown with blue dots with error bars\cite{Blanton08,Springob05}.
\label{fig:galaxyprop}
}
\end{figure*}

\begin{figure*}
\centering
\includegraphics[width=8cm]{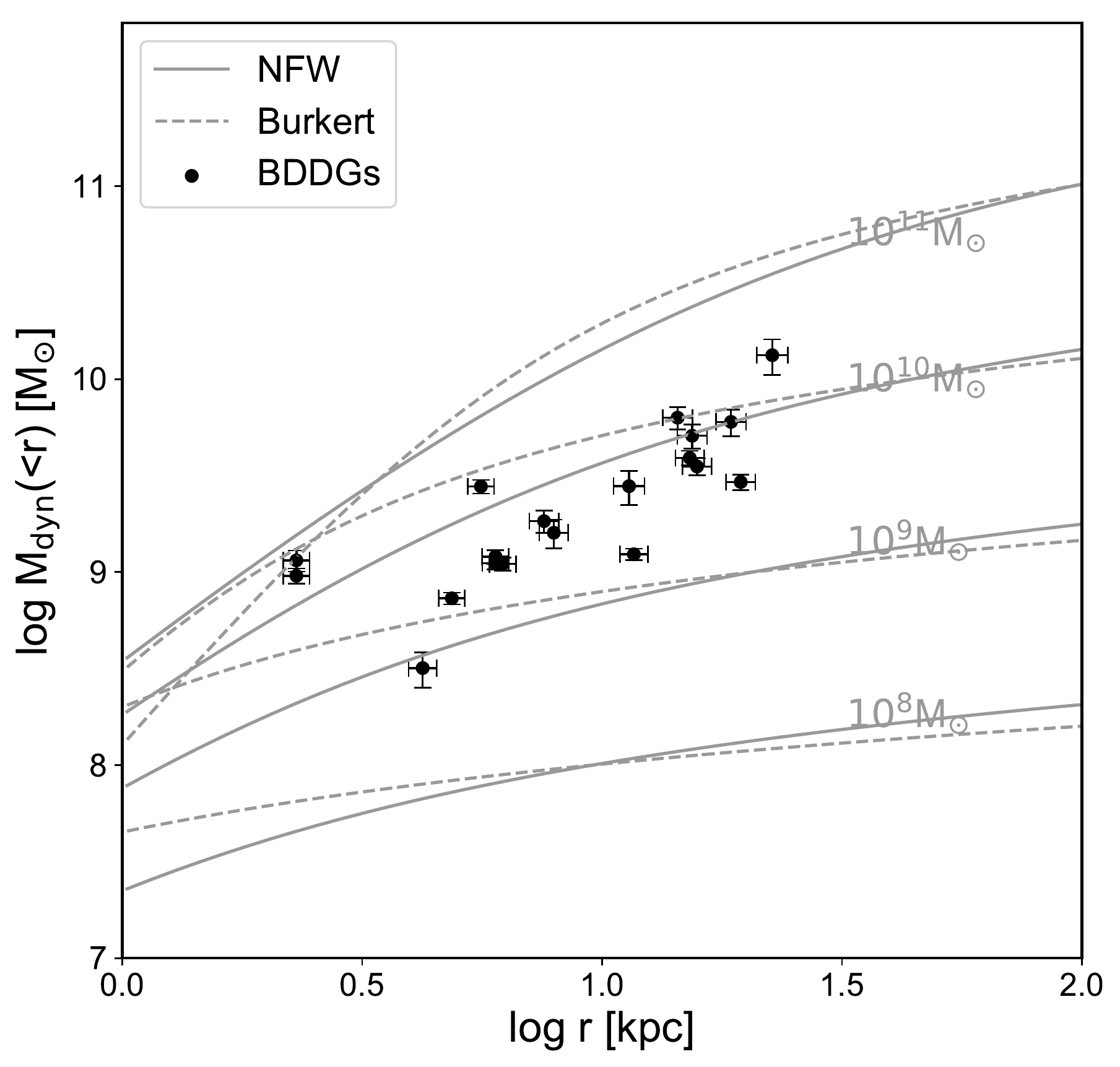}
\includegraphics[width=8cm]{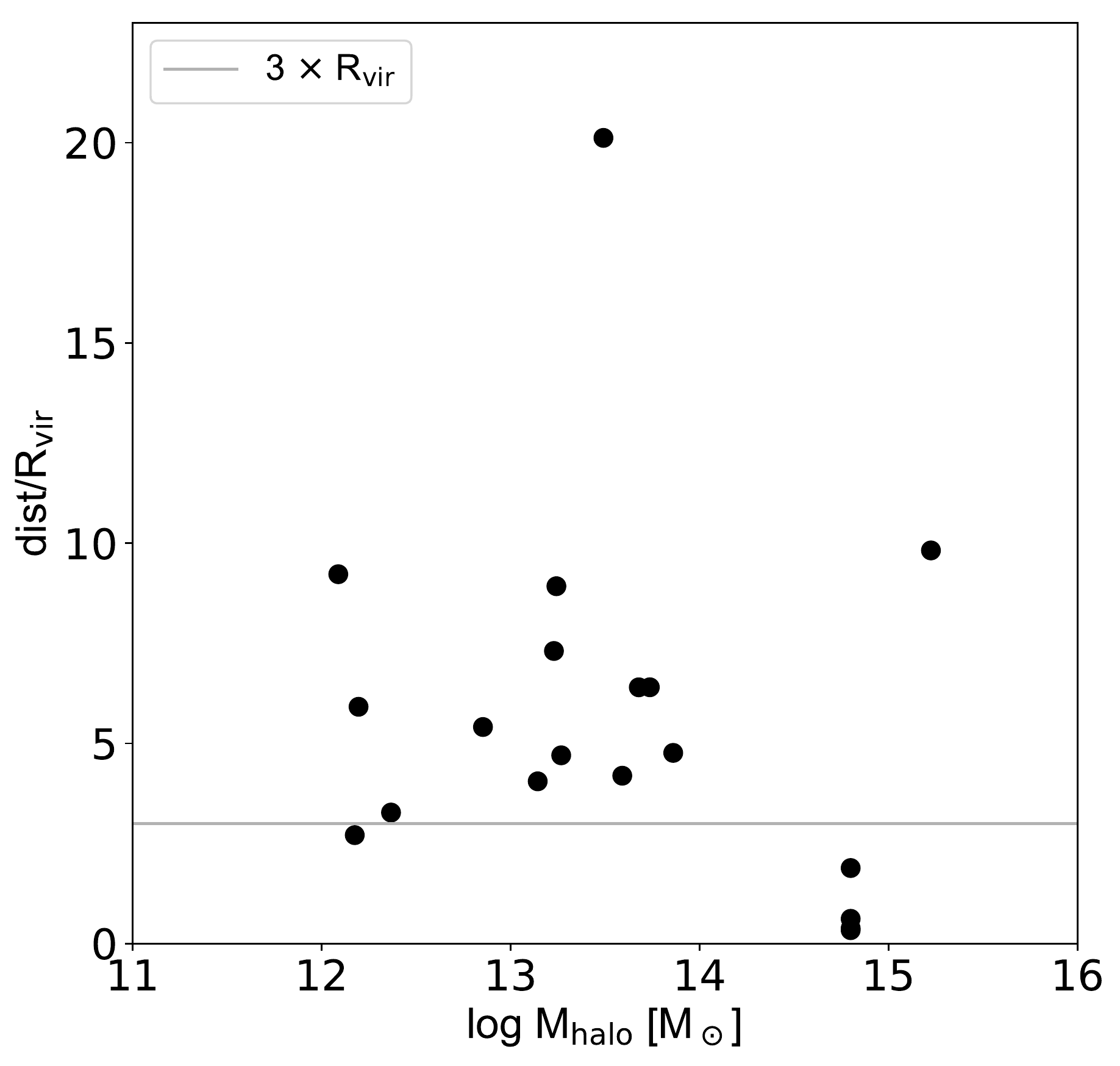}
\caption{
{\bf Halo mass and distance to the nearest clusters/groups.
Left panel}: Dynamical mass enclosed within r$_{\rm HI}$ vs. r$_{\rm HI}$ for baryon dominated galaxies. Black symbols are our 19 BDDGs. Solid and dashed curves are the typical dark halo mass profiles estimated assuming the NFW profile and the Burkert profile, ranking from $10^{11}$, $10^{10}$, $10^9$, to $10^8$ M$_{\odot}$ from top to bottom, respectively.
{\bf Right panel:} Distance of BDDGs to the nearest clusters/groups. The distance of each BDDG to its clusters/groups are normalized by the corresponding virial radius of each cluster/group. Solid horizontal line indicate the 3 $\times$ R$_{\rm vir}$ value, a limit where the effect of cluster/group could persists. We have 14 out of the 19 BDDGs beyond 3 $\times$ R$_{\rm vir}$ of their surrounding clusters/groups. Among the remaining 5 BDDGs, 4 are close to the Virgo cluster. 
\label{fig:halomass}
}
\end{figure*}

\clearpage

\pagestyle{empty}
\renewcommand{\thesubsection}{\Alph{subsection}}
\renewcommand{\thefigure}{\thesubsection\arabic{figure}}
\renewcommand{\thetable}{\thesubsection\arabic{table}}
\setcounter{subsection}{19}
\setcounter{figure}{0}
\setcounter{table}{0}
\section*{METHODS}
\subsection{ALFALFA.}
The Arecibo Legacy Fast ALFA (Arecibo L-band Feed Array) survey(ALFALFA)\cite{Giovanelli05a,Giovanelli07,Haynes07} is a blind extragalactic HI survey, covering a wide area of 7000 deg$^2$ of the sky at high Galactic latitudes. The beam size is about 3.5 arcmin. It has HI line spectra of more than 30,000 extragalactic sources with radial velocities $<$ 18000 km s$^{-1}$. The spectral resolution\cite{Giovanelli05b} is $\sim$10 km s$^{-1}$, enabling reliable measurements of $\sim$ 20 km s$^{-1}$, the typical velocity of dwarf galaxies. Here we use $\alpha$.40\cite{Haynes11}, a catalog which covers 40$\%$ of the full ALFALFA survey area, including the northern hemisphere spring region: 07$^{h}$30$^{m} <$ ra $<$ 16$^{h}$30$^{m}$, +04$^{\circ} <$ dec $<$ +16$^{\circ}$, and +24$^{\circ} <$ dec $<$ +28$^{\circ}$, and the northern hemisphere autumn region: 22$^{h}$00$^{m} <$ ra $<$ 03$^{h}$00$^{m}$, +14$^{\circ} <$ dec $<$ +16$^{\circ}$, and +24$^{\circ} <$ dec $<$ +32$^{\circ}$. There are 15855 HI detected sources, among which 15041 are extragalactic objects and the rest 814 are more likely Galactic high velocity cloud (HVC). 
 
The catalogue provides w$_{50}$, 50\% peak width of HI spectrum, and the corresponding errors. However, we find that in most cases, the edge of neither side of the spectrum has a cut-off feature. Instead, the flux at each side drops gradually. Calculations based on w$_{50}$ could lead to underestimates of the dynamical masses enclosed in r$_{\rm HI}$. w$_{20}$, 20\% peak width of the HI spectrum, is a better indicator of the gas velocity. In practice, we treat systems with one peak and two peaks in their HI spectra differently. For those with one peak, wavelengths at which the fluxes are 20\% of the peak values are identified to determine the w$_{\rm 20, raw}$, while for those with two peaks, the 20\%-fluxes are calculated for both peak values and the corresponding characteristic wavelengths are identified at each rising side of the spectrum to determine the w$_{\rm 20, raw}$. We perform Gaussian/Lorentzian fitting at each rising side. The results are consistent with those using polynomial fits. For each source, we generate 10000 realizations using the {\it rms} given by the ALFALFA catalogue, and adopt the corresponding 1-$\sigma$ range as the uncertainty in $\rm w_{\rm 20}$( $\rm w_{\rm 20,err}$). Then, we subtract the instrumental broadening as following\cite{Catinella12}

\begin{equation}
w_{20} = \sqrt[]{w^2_{20,raw} - (\delta s)^2}
\end{equation}
where $\delta s$ is the final velocity resolution after smoothing and we adopted the value of 10 km/s.

\subsection{SDSS DR7.}
The optical galaxy catalogue is based on SDSS DR7\cite{Abazajian09}, but with an independent set of improvement in sky-background subtraction. SDSS DR7 contains images and spectra for millions of galaxies, centered at z $\sim$ 0.1. The photometric data completes at 24 r-band apparent magnitude and the extinction-corrected spectroscopy data are complete at 17.72.

It is found that the SDSS photometric pipeline systematically overestimated the sky backgrounds of galaxies with extended low surface brightness by $\sim$ 0.5 mag\cite{He2013,Liu2008,Lisker2007}. The surface brightness of dwarf galaxies is usually very low and their luminosities could thus be very sensitive to the sky subtraction. Here we select a sub-sample of low surface brightness galaxy (central surface brightness in the B band $\rm \mu_{\rm 0}$(B)$\rm >22.5$ mag arcsec$\rm ^{\rm -2}$)\cite{Du15} and use the g-band and r-band SDSS mosaic images from the Twelfth Data Releases of the Sloan Digital Sky Survey(SDSS DR12)\cite{Alam15} to re-calculate their photometry\cite{Zheng2015}. The method is summarized as following: we first use SEXtractor\cite{Bertin1996} to extract and mask all sources above $2\sigma$ local background variation using the r-band/g-band image and then unmask the target galaxy by visual inspection. We grow the masks using the IDL function DILATE with a square shape operator, whose size is proportional to the size of the mask. Stars on top of our target galaxies are removed and filled with the average value of pixels around the star. We fit each galaxy's ellipticity profile using the ELLIPSE task in IRAF and use the ellipticity of the largest fitted ellipse to calculate the r-band axis ratio b/a. The Petrosian flux is then calculated by summing up all fluxes within 2 times the elliptical Petrosian radius\cite{Blanton2001,Yasuda2001}. For those with smaller surface brightness, we simply adopt the Petrosian magnitudes, and use exponential axes ratio expAB$_{\rm r}$ (or de Vaucouleurs axes ratio deVAB$_{\rm r}$) if the likelihood fraction of de Vaucouleurs model fracDeV$_{\rm r} <$ 0.5 (fracDeV$_{\rm r} >$ 0.5)  as the b/a value from the released SDSS DR7 catalogue.

There are 12423 galaxies with optical counterparts (OCs) in the HI and optical cross-matched catalogue\cite{Haynes11}. Here we focus on dwarf galaxies with absolute magnitude M$_{\rm r}>-18$ and high signal-to-noise ratio in HI spectra: SNR$_{\rm HI}>10$. To avoid significant bias caused by inclination, we further require the r-band b/a to be within 0.3 and 0.6, corresponding to the inclination angle range of [54.7, 76.8] degree. The final parent sample consists of 324 galaxies.

\subsection{LITTLE THINGS.}
``Local Irregulars That Trace Luminosity Extremes, The HI Nearby Galaxy Survey"(LITTLE THINGS)\cite{Hunter12} is a high resolution, very large array HI survey with angular resolution of $\sim$6 arcsec and velocity resolution of $<$2.6 km s$^{-1}$, aiming to obtain deep HI-line maps of dwarf galaxies. The high resolution enable a reliable measurement of the rotation curve\cite{Oh15}. In practice, Oh et al.\cite{Oh15} fitted a 2D tilted-ring model to the 2D bulk velocity fields extracted from the HI data. This provides geometric and kinematic parameters for each galaxy, including central position, position angle, inclination, systemic velocity and rotation velocity. The asymmetric drift is then made to improve the rotation curve measurement by taking into account the dynamical support due to random motions. In combination with Spitzer 3.6 $\mu$m and optical U, B, and V images, the rotation curve is then used to analyze the dynamical contributions by baryons and dark matter. They found the inner slope of the dark matter density profiles deviate significantly from the cusp-like mass distribution predicted by dark matter only simulations, but in line with those predicted by $\rm \Lambda$CDM + hydro-dynamical simulations, for which the baryon feedback process could change the inner profile.

\subsection{Mass algorithm.}
Four kinds of masses are used in this work, including stellar mass, HI mass, dynamical mass, and dark halo virial mass. Here we describe briefly how each of the mass is derived.

Gas mass. Gas in galaxies mainly consists of HI and Helium. The HI mass is calculated following Roberts (1962)\cite{Roberts62}

\begin{equation}
{\rm M_{HI}} = 2.36\times 10^{5}d^{2}\int {\rm F} \mathrm{d} {\rm v [M_{\odot}]}
\end{equation}
where F is the HI flux in Jy km s$^{-1}$, and v is the receding velocity. d is the distance\cite{Haynes11} of the HI detected galaxies estimated using d = cz/H$_0$ for galaxies with cz$>$6000 km/s, where H$_{0}$ is the Hubble constant, 70km s$^{-1}$, and c is the speed of light. For galaxies with cz$<$6000 km/s, the distance is derived adopting the local universe peculiar velocity model\cite{Masters05}, using a combination of primary distances form the literature and secondary distances from the Tully-Fisher relation. 
The total gas mass is derived by assuming the same Helium-HI ratio as that derived from the Big Bang nucleosynthesis: M$_{\rm gas} = 1.33 \times$M$_{\rm HI}$ (Y$_{\rm P}^{\rm BBN}=0.247$)\cite{Planck18}, where Y$_{\rm P}^{\rm BBN}$ is the primordial He abundance from Planck in $\Lambda$CDM.

Stellar mass. 
For a given galaxy, we calculate the absolute magnitude in g and r band, using the photometry as described above and the distance from $\alpha.$ 40. Following Bell's algorithm, we have 

\begin{equation}
\log_{10}({\rm M/L}) = a_\lambda +b_\lambda \times ({\rm M}_g - {\rm M}_r).
\end{equation}
Here we adopt a Kroupa initial mass function(IMF)\cite{Kroupa03} with $a_{\rm r} = -0.306 -0.15$ and $b_{\rm r} = 1.097$. The final error bars accounts for both observational errors in photometry and distance, and the method error of 0.1 dex as suggested by Bell\cite{Bell03}.

Dynamical mass. 
With the velocity and radius of the HI component, we calculate the dynamical mass enclosed in r$_{\rm HI}$ by assuming a spherical symmetry:
M$_{\rm dyn}$ ($<$r$_{\rm HI}$) =V$^2_{\rm HI}$r$_{\rm HI}$/G, where G is the gravitational constant. Previous work\cite{Wang16} found that there is a tight relation between the HI mass and the HI radius r$_{\rm HI}$, defined at a surface density of 1M$_\odot$ pc$^{-2}$.

\begin{equation}
\rm log_{10} r_{\rm HI}=(0.506 \pm 0.003) log_{10} {\rm M_{HI}}  -  (3.293 \pm 0.009) - log_{10} 2
\end{equation} 
 This relation  has been proven successful in describing the r$_{\rm HI}$ and M$_{\rm HI}$ relation for various galaxies types\citep{Wang16}. In particular,  we apply this r$_{\rm HI}$ vs. M$_{\rm HI}$ relation to the dwarf galaxies from LITTLE THINGS (Left panel of Extended Data Figure 1) and find they agree very well with the direct measurement using the resolved HI images.

The velocity V$_{\rm HI}$ is estimated using w$_{20}$: V$_{\rm HI}$ = w$_{20}$/2/sin(i), where i is the inclination angel.

\begin{equation}
{\rm sin(i)} = \sqrt[]{\frac{1-(\frac{b}{a})^2}{1-q_0^2}}
\end{equation}
The parameter q$_0$ is the intrinsic axial ratio of a galaxy seen edge-on. The value of q$_0$ could depend on galaxy morphology\cite{Giovanelli97,Tully09}. Here we use 0.2 as usually adopted in the literature. w$_{20}$ is a good indicator of the circular velocity given sharp edges at both sides of the given spectrum. However, for dwarf galaxies, velocity dispersion could contribute a non-negligible fraction to the total kinematic energy of the systems. This is reflected by the shape of the spectrum that at both sides the spectrum decline gradually, lacking the feature of a sharp cut-off. w$_{20}$ could thus carry both information of the circular velocity and of the velocity dispersion. We test the choice of using w$_{20}$ as an indicator of the asymmetric drift-corrected circular velocity using dwarf galaxies in the LITTLE THINGS (Middle panel of Extended Data Figure 1). It shows the w$_{20}$ is a good indicator of the corrected circular velocity, while w$_{50}$ usually under-estimates the corrected circular velocity.
We then test the full method by performing the same analysis as for our BDDGs on the LITTLE THINGS galaxies and find the derived  M$_{\rm dyn}$ are consistent with the estimates using the tilted-ring model (Right panel in Extended Data Figure 1).

Note for we do not have HI images, we use the inclination angels estimated in the optical r-band. There could be mis-alignment between r-band image and HI velocity field, leading to biased estimate of the circular velocity. We restrict our parent sample to have r-band b/a between [0.3, 0.6] in order to avoid severe deviation of the derived velocities from the true ones.  Using dwarf galaxies in the LITTLE THINGS, we show in the left panel of Extended Data Figure 2 the distribution of the angle between r-band image and the HI velocity filed. It shows that in most cases the mis-alignment is below 20 degree, thought there is a tail towards large angle. We apply the same distribution of the mis-alignment to our parent sample and generate 10000 realization for each parent galaxy. The resulting M$_{\rm dyn}$/M$_{\rm bary}$ distribution is shown in the right panel of Extended Data Figure 2. It shows there is still an apparent excess of BDDGs at the low M$_{\rm dyn}$/M$_{\rm bary}$ end. The existence of BDDGs is also supported when applying the same distribution of the mis-alignment as those in the simulations\cite{Starkenburg2019}.

If the BDDGs were very compact dwarf galaxies, it would be possible to have baryon mass dominated in the compact region. We compare the r-band r$_e$ and r$_{\rm HI}$ of the BDDGs and those of the parent dwarf galaxies and find that the BDDGs are slightly more extended than the parent dwarf galaxies, both in the optical and in the HI contents (Extended Data Figure 3). In addition, r$_{\rm HI}$ is larger than r$_{\rm e}$ for all the BDDGs. Both the baryonic masses and dynamical masses are measured at radii within which it is expected to be dominated by dark matter if the BDDGs were typical dwarf galaxies.

Halo mass. Assuming a typical NFW profile\cite{NFW97}, r$_{\rm HI}$ and M$_{\rm dyn}$($<$r$_{\rm HI})$ are related as

\begin{equation}
\begin{aligned}
{\rm M_{dyn}(<r_{HI})}& =  \int_0^{r_{\rm HI}}4\pi r^2\rho(r)\mathrm{d}r
 \\
 & =  4\pi \rho_0 R_s^3 \left [\ln \left(\frac{R_s +r_{\rm HI}}{R_s} \right)-\frac{r_{\rm HI}}{R_s+r_{\rm HI}}\right],
 \end{aligned}
 \end{equation}
where R$_{\rm s}$ is a free parameter. Given r$_{\rm HI}$ and M$_{\rm dyn}$($<$r$_{\rm HI})$, for each galaxy, we obtain the corresponding R$_{\rm s}$. The virial mass is expressed as

\begin{equation}
\rm{ M}_{vir}  =  \int_0^{R_{vir}}4\pi r^2\rho(r)\mathrm{d}r =  4\pi \rho_0 R_s^3 \left [\ln \left(1+c \right)-\frac{c}{1+c}\right], 
 \end{equation} 
 \begin{equation}
\rm \rho_0 = \frac{200 \rho_{cr}}{3}\frac{c^3}{\ln (1+c) - \frac{c}{1+c}},
\end{equation}
where $\rm \rho_0$ is a characteristic density, $\rm \rho_{cr}$ is the critical density, and c is the concentration parameter which relates to R$_s$ and R$_{\rm vir}$ via c = $\frac{\rm R_{\rm vir}}{\rm R_s}$. The concentration parameter follows a tight halo mass-concentration relation\cite{Dutton14}.

\begin{equation}
 \rm  log_{10} c = 0.905 - 0.101({\rm log_{10} M}_{vir}/10^{12}h^{-1}M_{\odot}).
\end{equation}
Here h = H$_0$/(100 km s$^{-1}$ Mpc$^{-1}$)=0.7. Inserting Eq. 9 to Eq. 7, one could obtain the M$_{\rm vir}$. 

Assuming a Burkert profile\cite{Burkert95}, r$_{\rm HI}$ and M$_{\rm dyn}$($<$r$_{\rm HI})$ are related as

\begin{equation}
\begin{aligned}
{\rm M_{dyn}(<r_{HI})}& =  \int_0^{r_{\rm HI}}4\pi r^2\rho(r)\mathrm{d}r
 \\
 & =  2\pi \rho_0 r_0^3 \left [\ln \left(1+\frac{r_{\rm HI}}{r_0} \right)+\frac{1}{2}\ln \left(1+\frac{r_{\rm HI}^2}{r_0^2} \right) -arctan(\frac{r_{\rm HI}}{r_0})\right],
 \end{aligned}
 \end{equation}
 where r$_0$ and $\rho_0$ are free parameters. With the relation between r$_0$ and M$_{\rm vir}$\cite{Salucci07}  

\begin{equation}
  {\rm log_{10}} (r_0/{\rm kpc}) = 0.66 - 0.58({\rm log_{10} M}_{\rm vir}/10^{11}M_{\odot}),
\end{equation}
one could derive M$_{\rm vir}$ for any given combination of r$_{\rm HI}$ and M$_{\rm dyn}$($<$r$_{\rm HI})$.
\subsection{Galaxy clusters.}
The galaxies groups and clusters catalogues are created based on the SDSS DR12\cite{Alam15} and the 2MASS Redshift Survey (2MRS)\cite{Huchra12}, using a friends-of-friends group finder algorithm\cite{Saulder16}. Various observational biases have been taken into account, including Malmquist bias, the `Fingers of God' and etc. The groups/clusters mass depends on the total luminosity, the luminosity distance, the velocity dispersion, the group radius and the number of detected group members. This method has been carefully calibrated and optimized on wide-angle mock catalogues from the Millennium simulation\cite{Guo11,Guo13}. For any group/cluster included in these catalogues, where available, group parameters in the published literature are adopted primarily. 

The properties of the 19 BDDGs are summarized in Extend Data Table 1. 
\\

\bibliographystyle{plainnat}

\smallskip
\noindent {\small {\bf Acknowledgements} \\
This work is supported by the National Key R\&D Program of China (Nos 2018YFA0404503, 2016YFA0400703 and 2016YFA0400702), the National Natural Science Foundation of China (No.11573033, 11622325, 11133003, 11425312, 11773001, 11721303, 11733006, 11703036) and the National Natural Science Foundation of
China (grant numbers 11573033, 11622325, 11133003, 11425312, 11773001, 11721303, 11733006, 11703036 and
U1931110). Q.G. and L.G. acknowledge support from the Royal Society Newton Advanced Fellowships. Z.Z acknowledges the support by the Open Project Program of the Key Laboratory of FAST, Chinese Academy of Sciences.
}

\smallskip
\noindent {\small {\bf Author contributions} \\
 Q.G. led and played a part in all aspects of the analysis, and wrote the manuscript. H.H compiled the data and carried out most of the data reduction and analysis. Z.Z. measured the optical photometry of low-surface brightness galaxy. S.L carried out most of the analysis related to various mass determinations. All authors contributed to the analysis,
and to the writing of the manuscript.
}

\smallskip
\noindent {\small {\bf Additional information} \\
Supplementary information is available in the on-line version of the paper.
}

\smallskip
\noindent {\small {\bf Competing financial interests} \\
 The authors declare that they have no competing financial interests. Correspondence and requests for materials should be addressed to Q.G. (guoqi@nao.cas.cn).

}

\smallskip
\noindent{\small{\bf Data availability.} The ALFALFA data are available in
the The Arecibo Legacy Fast ALFA Survey (http://egg.astro.cornell.edu/alfalfa/data/index.php) and the SDSS data are available at the Sloan Digital Sky Survey (https://www.sdss.org/) . The other
data that support the results of this study are available from
the corresponding author upon reasonable request.}

\smallskip
\noindent{\small{\bf Code availability.} We use  standard data
reduction tools in the IDL and Python environments, and
the publicly available codes SExtractor (https://www.astromatic.net/software/sextractor) for this study.}

\newpage
\appendix

\renewcommand{\thefigure}{ Figure \arabic{figure}}   
\renewcommand{\thetable}{Table \arabic{table}} 
\renewcommand{\figurename}{Extended Data}
\renewcommand{\tablename}{Extended Data}
\begin{landscape}
\begin{table}
\centering
\newsavebox{\tablebox}
\begin{lrbox}{\tablebox}
\begin{threeparttable}
 \begin{tabular}{llllllllllllllll}
    \hline \hline
    {\textbf{AGCNr}} & {\textbf{RA}} & {\textbf{DEC}} &  {\textbf{distance}} & {\textbf{V$_{\rm \textbf{helio}}$}} & 
    {\textbf{w20}} & {\textbf{w20er}} & {\textbf{M$_{\rm \textbf{g}}$}} &  {\textbf{M$_{\rm \textbf{r}}$}} & {\textbf{re}} & {\textbf{log M$_{\rm \textbf{HI}}$}} & 
    {\textbf{log M$_{\rm \textbf{stellar}}$}} & {\textbf{log M$_{\rm \textbf{dy}}$}} & {\textbf{log M$_{\rm \textbf{vir}}$}} & {\textbf{b/a ratio}} & {\textbf{dist$_{\rm \textbf{gr}}$/R$_{\rm \textbf{vir,gr}}$}}        
    \\
       & [deg (\degree)] & [deg (\degree)] & [Mpc] & [km/s] & 
    [km/s] & [km/s]& [mag] & [mag] & [kpc] & [M$_{\rm \odot}$] & 
     [M$_{\rm \odot}$] &  [M$_{\rm \odot}$] & [M$_{\rm \odot}$] &  
     &
 \\
\hline
\textbf{AGC 6438} & 171.47292 & 9.98695 & 20.4 & 1156 & 80.36 & 2.03 & -17.34 & -17.75 & 0.59 & 8.58 & 8.951 & 9.444 & 10.231 & 0.524 & 2.707
\\
\textbf{AGC 6980$^{*}$} & 179.77542 & 24.47278 & 50.3 & 3342 & 56.63 & 1.54 & -17.74 & -17.89 & 4.93 & 9.44 & 8.731 & 9.592 & 9.876 & 0.55 & 9.224
\\
\textbf{AGC 7817} & 189.76666 & 13.36306 & 16.7 & 2737 & 82.37 & 4.45 & -16.77 & -17.2 & 0.84 & 7.82 & 8.749 & 9.061 & 10.599 & 0.491 & 0.334
\\
\textbf{AGC 7920} & 191.18918 & 12.35028 & 16.7 & 849 & 79.03 & 2.6 & -17.2 & -17.71 & 1.54 & 7.82 & 9.046 & 8.981 & 10.653 & 0.4 & 0.617
\\
\textbf{AGC 7983} & 192.44583 & 3.84222 & 16.6 & 694 & 46.12 & 0.83 & -15.01 & -15.21 & 0.77 & 8.64 & 7.702 & 9.046 & 9.515 & 0.6 & 1.885
\\
\textbf{AGC 9500} & 221.33916 & 7.8625 & 27.6 & 1690 & 39.08 & 0.31 & -17.34 & -17.56 & 5.19 & 9.21 & 8.667 & 9.092 & 9.712 & 0.444 & 4.761
\\
\textbf{AGC 191707} & 144.44835 & 27.56833 & 25.1 & 1595 & 49.27 & 1.21 & -16.0 & -16.31 & 1.64 & 8.64 & 8.273 & 9.08 & 9.567 & 0.57 & 5.408
\\
\textbf{AGC 205215} & 163.98709 & 13.97361 & 106.7 & 7125 & 72.5 & 4.41 & -17.03 & -17.33 & 2.86 & 9.45 & 8.66 & 9.706 & 9.984 & 0.334 & 9.818
\\
\textbf{AGC 213086$^{\dag}$} & 171.80333 & 8.78806 & 92.8 & 6145 & 78.35 & 4.33 & -17.36 & -17.62 & 3.39 & 9.39 & 8.741 & 9.8 & 10.149 & 0.469 & 7.306
\\
\textbf{AGC 220901} & 189.99374 & 13.78139 & 17 & 1005 & 45.38 & 0.74 & -14.18 & -14.07 & 0.74 & 8.46 & 6.915 & 8.864 & 9.363 & 0.486 & 0.38
\\
\textbf{AGC 241266$^{\dag}$} & 213.8425 & 14.24972 & 77.4 & 5249 & 52.82 & 1.98 & -17.27 & -17.75 & 4.61 & 9.47 & 9.02 & 9.547 & 9.96 & 0.551 & 6.4
\\
\textbf{AGC 242440} & 212.19499 & 4.90972 & 76.9 & 5260 & 42.47 & 1.18 & -17.52 & -18.0 & 0.72 & 9.65 & 9.133 & 9.467 & 10.098 & 0.576 & 8.924
\\
\textbf{AGC 258421} & 233.61292 & 6.28583 & 145.4 & 10034 & 87.79 & 8.53 & -17.79 & -17.95 & 2.88 & 9.78 & 8.76 & 10.124 & 10.387 & 0.518 & 4.701
\\
\textbf{AGC 321435} & 341.93207 & 32.18834 & 56.6 & 3863 & 56.83 & 4.41 & -16.72 & -16.73 & 2.17 & 8.88 & 8.105 & 9.204 & 9.593 & 0.326 & 20.122
\\
\textbf{AGC 331776} & 351.52747 & 14.26417 & 41.2 & 2860 & 29.59 & 2.9 & -11.77 & -11.83 & 0.22 & 8.34 & 6.196 & 8.503 & 8.904 & 0.592 & 6.399
\\
\textbf{AGC 733302} & 224.41417 & 26.66472 & 23.0 & 1280 & 48.36 & 0.99 & -16.2 & -16.47 & 0.39 & 8.67 & 8.276 & 9.042 & 9.489 & 0.515 & 5.913
\\
\textbf{AGC 749244$^{\dag}$} & 190.81416 & 27.26361 & 112.7 & 7608 & 70.87 & 4.91 & -15.29 & -14.99 & 1.29 & 9.61 & 7.066 & 9.778 & 10.003 & 0.365 & 3.271
\\
\textbf{AGC 749445$^{\dag}$} & 179.17708 & 24.96306 & 64.8 & 4374 & 54.51 & 3.06 & -15.57 & -15.72 & 1.3 & 8.84 & 7.863 & 9.264 & 9.708 & 0.563 & 4.191
\\
\textbf{AGC 749457} & 186.755 & 25.16611 & 101.9 & 6840 & 58.68 & 5.49 & -16.74 & -17.01 & 2.13 & 9.19 & 8.504 & 9.445 & 9.759 & 0.464 & 4.051
\\
\hline
\\

 \end{tabular}
  \begin{tablenotes}
        \footnotesize
        \item[*]In the literature AGC 6980 was defined as Abell 1367's member(Leo cluster: $\rm cz = 6459$, ra = 176.15083, dec=19.77194 )\cite{Rines03}. The projected distance between the AGC 6980 and Abell 1367 is about 8 Mpc, well beyond 3 R$_{\rm vir}$ of the Abell 1367. 
        \item[$ ^\dag$]The minimum projected distance of the BDDG to its surroundings clusters with relatively small difference in redshifts (cz$_{\rm BDDG}$ - cz$_{\rm cluster}$ $<$ 10$\times$R$_{\rm cluster,vir}$) is $<$ 3$\times$R$_{\rm vir}$ .   
      \end{tablenotes}
\end{threeparttable}
\end{lrbox}
\caption{{\bf BDDGs data}. Column 1: ALFALFA galaxy ID; Cols. 2and 3: equatorial coordinates; Col. 4: distance judged by ALFALFA group; Col. 5 heliocentric velocity; Col. 6: velocity width 20\% of the peak flux; Col. 7: error of w$_{\rm 20}$; Cols. 8 and 9: g-band, r-band magnitude; Col. 10: effective radius in r-band; Col. 11: neutral hydrogen mass; Col. 12: stellar mass; Col.13: dynamical mass within r$_{\rm HI}$; Col. 14: virial mass; Col. 15: r-band minor-to-major axis ratio; Col. 15: distance to the nearest group over virial radius}
\scalebox{0.75}{\usebox{\tablebox}}
\end{table}
\end{landscape}

\begin{figure*}
\centering
\includegraphics[width=17cm]{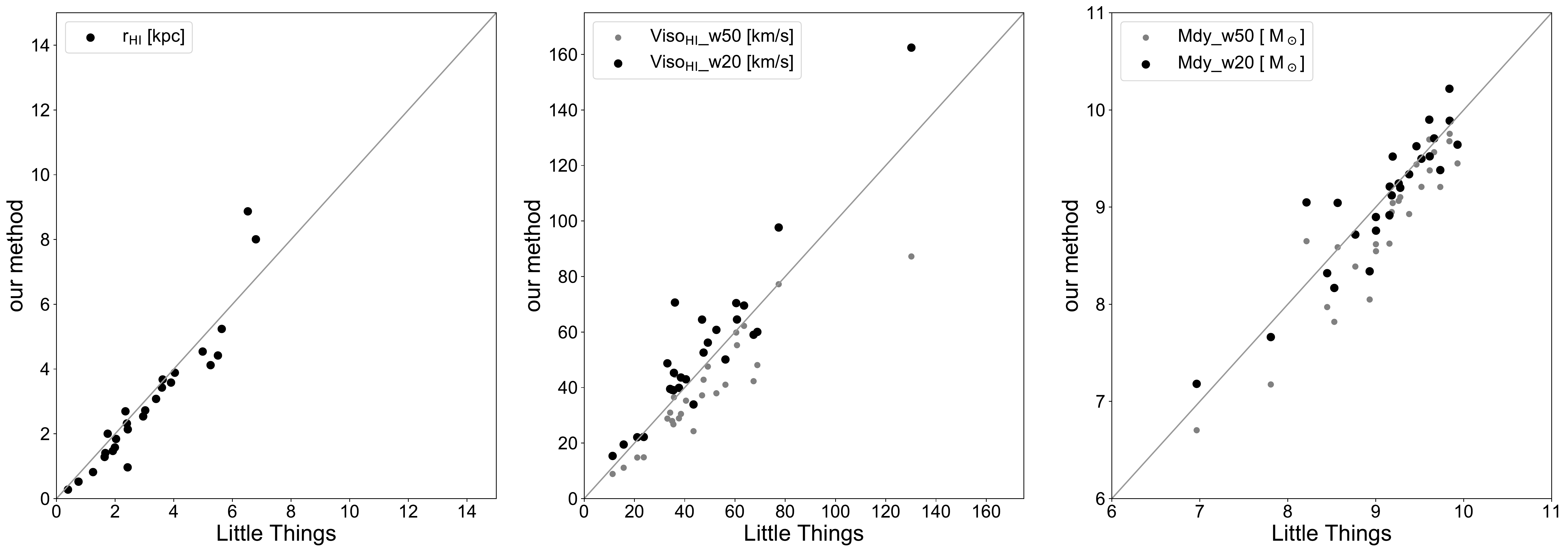}
\caption{
{\bf Comparison of the galaxy properties between  different methods for dwarf galaxies in LITTLE THINGS.}
{\bf Left panel:} Comparison of the r$_{\rm HI}$ using our method with that obtained using tilted-ring method by LITTLE THINGS team.{\bf Middle panel:} Comparison of the V$_{\rm HI}$ using our method with that obtained using tilted-ring method by LITTLE THINGS team. {\bf Right panel:} Comparison of the dynamical mass estimated using our method with those derived using the tilted-ring method. Black and grey dots are measurements based on w$_{20}$ and w$_{50}$, respectively. 
\label{fig:CompareLT}
}
\end{figure*}

\begin{figure*}
\centering
\includegraphics[width=5.2cm]{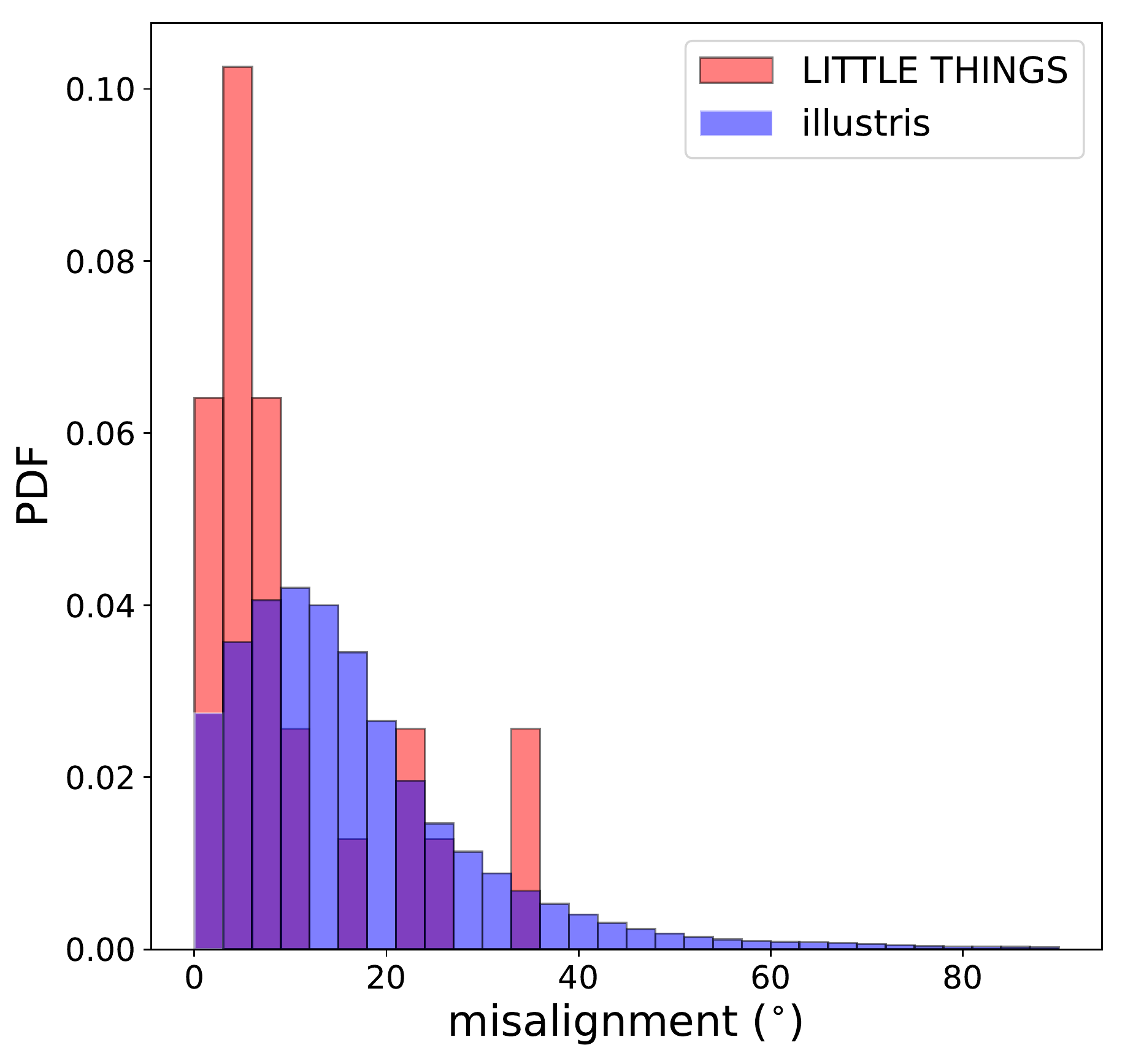}
\includegraphics[width=5.2cm]{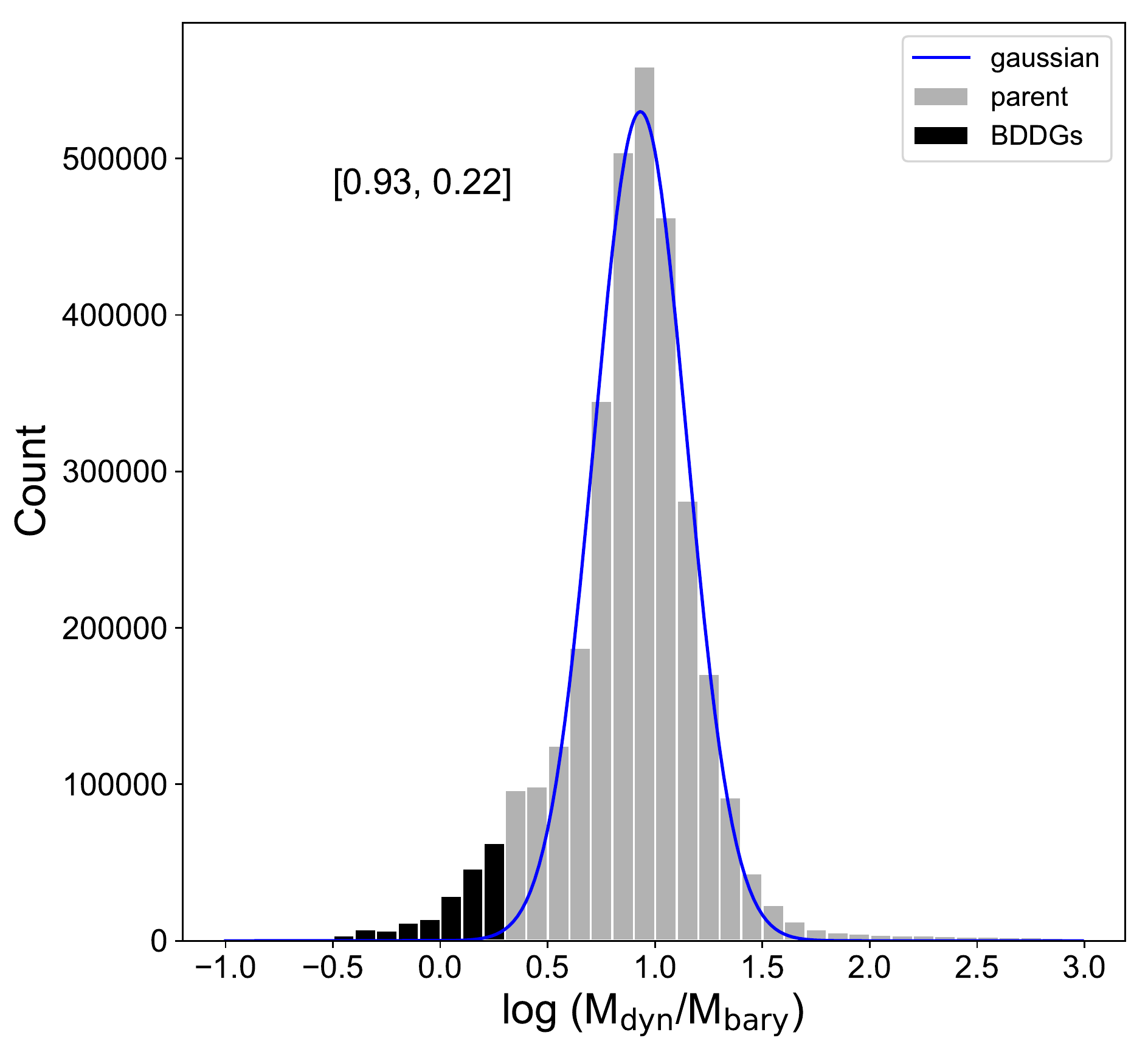}
\includegraphics[width=5.2cm]{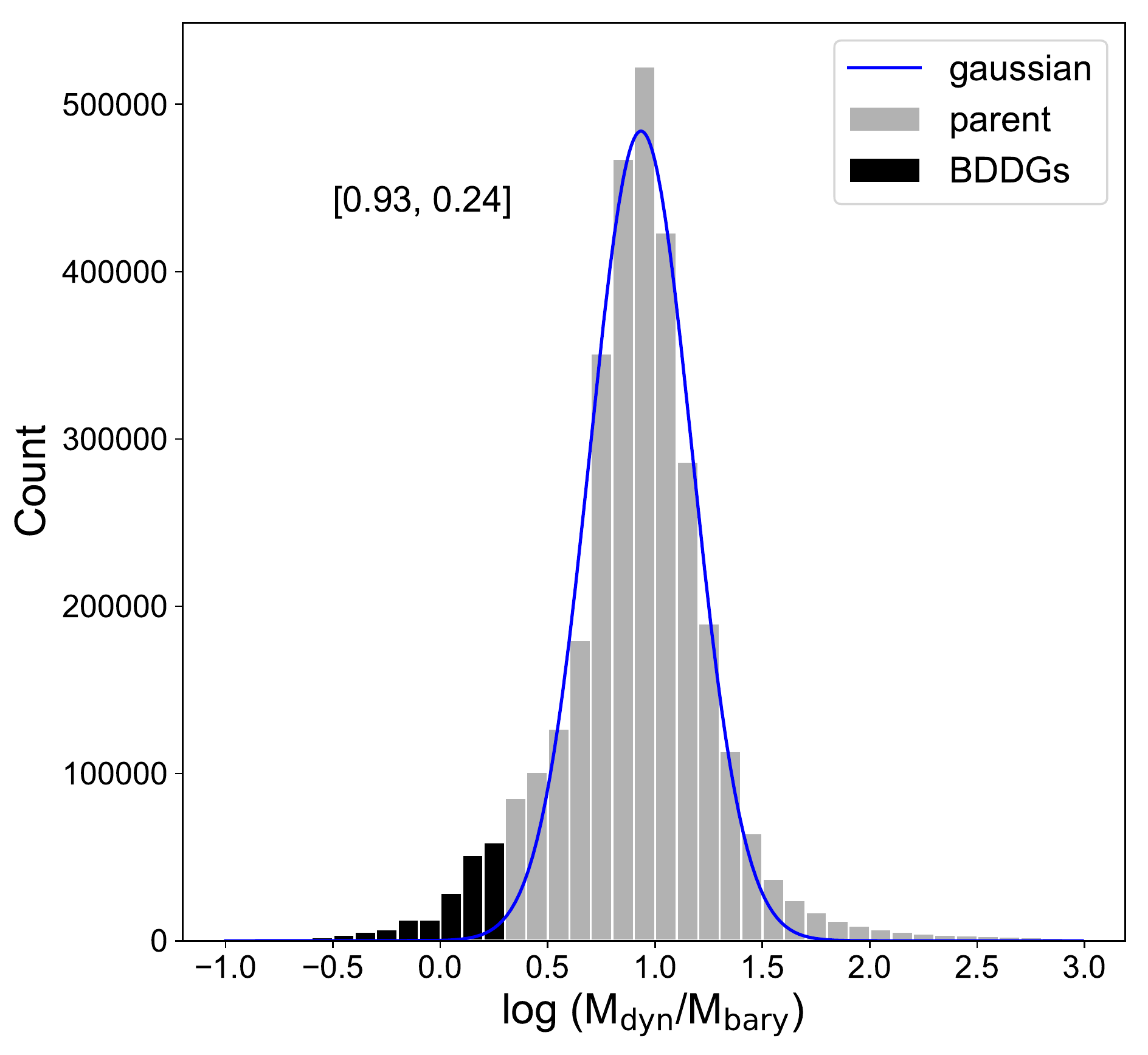}
\caption{{\bf Distribution of the M$_{\rm dyn}$-to-M$_{\rm bary}$ ratio with different misalignments between optical image and HI velocity field.} {\bf Left panel:} The distribution of the misalignment between optical image and HI velocity field for dwarf galaxies in LITTLE THINGS(red)/Illustris(blue). {\bf Middle/Right panel:} Distribution of the M$_{\rm dyn}$-to-M$_{\rm bary}$ ratio for our parent sample after applying the same distribution of misalignment between optical  image and HI velocity field from the LITTLE THINGS/Illustris in the left panel. Grey histogram is based on 10000 random realizations and blue curve is the Gaussian fit. Samples classified as BDDGs are high-lighted with black histogram.
\label{fig:Misalignment}
}
\end{figure*}

\begin{figure*}
\centering
\includegraphics[width=8cm]{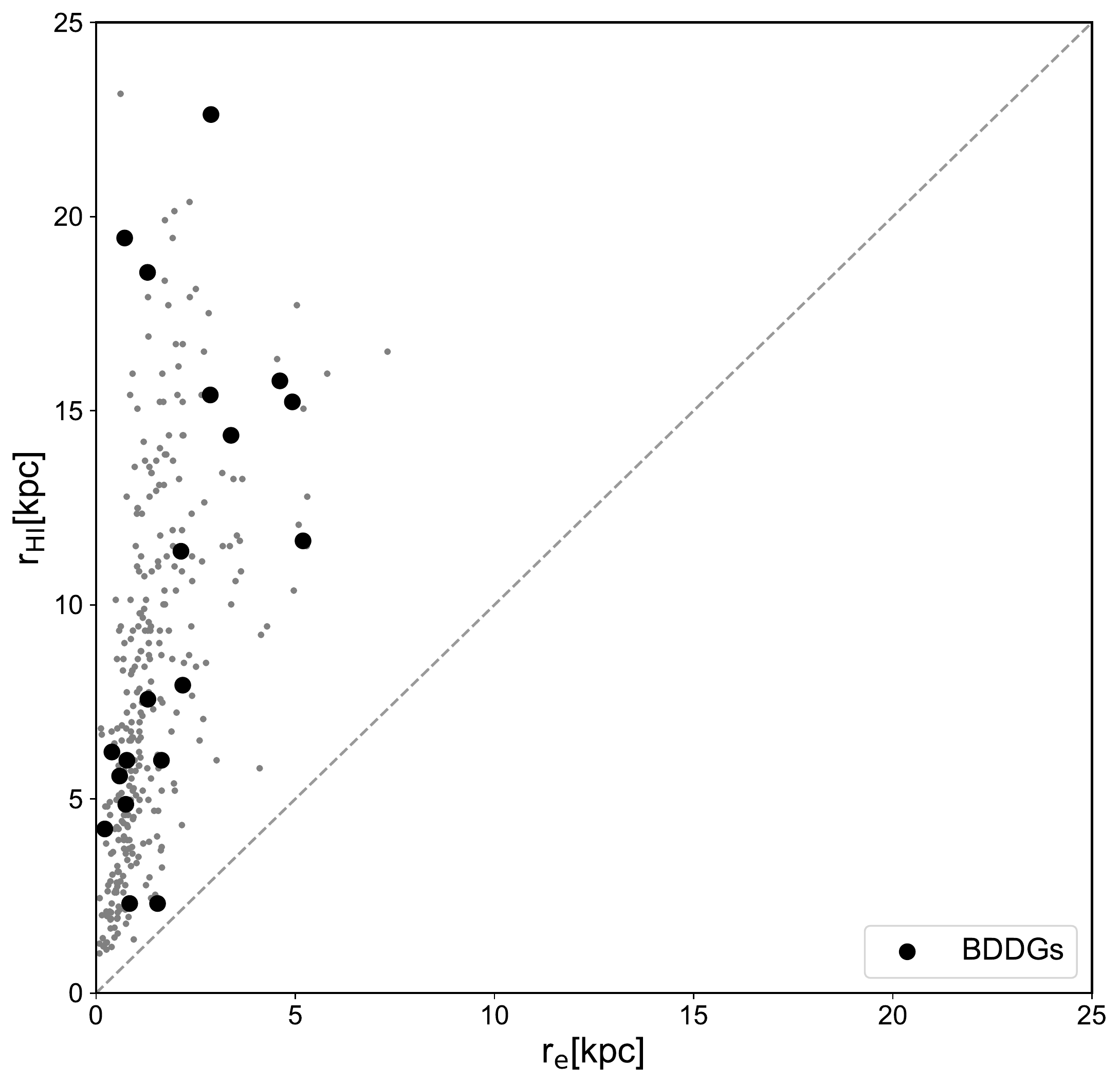}
\includegraphics[width=8cm]{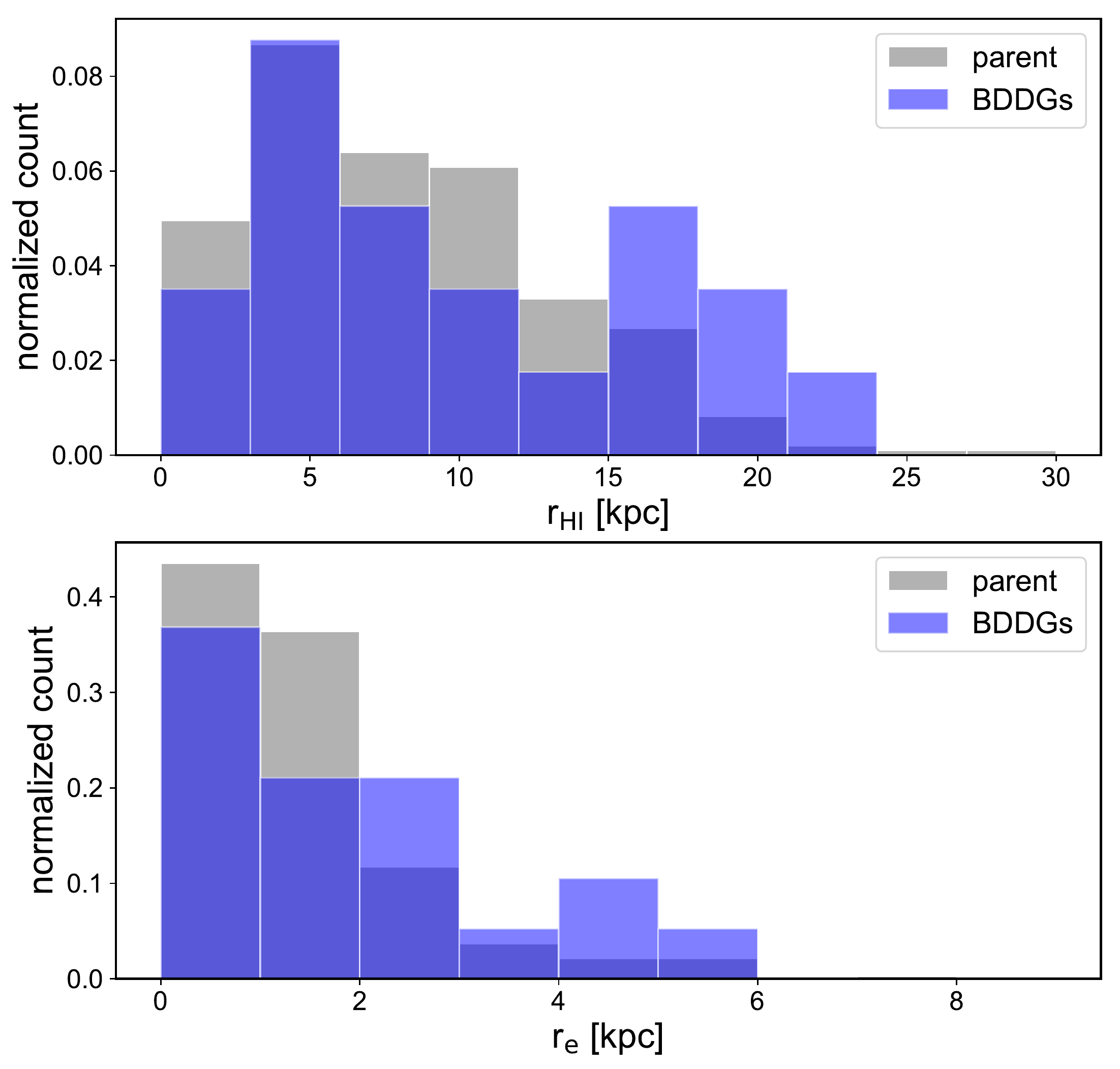}
\caption{{\bf Comparison of r$_{\rm HI}$ and r$_{\rm e}$ between the parent sample and the BDDGs.}
{\bf Left panel:} r$_{\rm HI}$ and r$_{\rm e}$ relation of the parent sample(grey dots) and the BDDGs(black dots).
{\bf Upper right panel:} the distribution of r$_{\rm HI}$, grey histogram and blue histogram represent parent sample and BDDGs respectively. 
{\bf Bottom right panel:} the distribution of r$_{\rm e}$, grey histogram and blue histogram represent parent sample and BDDGs respectively. 
\label{fig:Re_Rhi_distri}
}
\end{figure*}

\end{document}

%% file: preamble.tex
\usepackage{setspace}
\usepackage{amsfonts}
\usepackage{amssymb}
\usepackage{amsmath}

\usepackage{float}
\usepackage{placeins}
\usepackage{graphicx}

\usepackage[export]{adjustbox}

\usepackage[bf,bf,bf]{caption}

\renewcommand{\figurename}{Figure}
\renewcommand{\tablename}{Table}

\usepackage[super,comma,sort&compress]{natbib}

\usepackage{hyperref}
\usepackage{verbatim}

\input{preamble_optional.tex}

%% file: preamble_optional.tex
\usepackage{ulem} 
\usepackage{rotating} 
\usepackage{soul}

\usepackage{color} 

\usepackage[colorinlistoftodos]{todonotes}

\makeatletter
\newsavebox\myboxA
\newsavebox\myboxB
\newlength\mylenA

\newcommand*\xoverline[2][0.75]{%
    \sbox{\myboxA}{$\m@th#2$}%
    \setbox\myboxB\null
    \ht\myboxB=\ht\myboxA%
    \dp\myboxB=\dp\myboxA%
    \wd\myboxB=#1\wd\myboxA
    \sbox\myboxB{$\m@th\overline{\copy\myboxB}$}
    \setlength\mylenA{\the\wd\myboxA}
    \addtolength\mylenA{-\the\wd\myboxB}%
    \ifdim\wd\myboxB<\wd\myboxA%
       \rlap{\hskip 0.5\mylenA\usebox\myboxB}{\usebox\myboxA}%
    \else
        \hskip -0.5\mylenA\rlap{\usebox\myboxA}{\hskip 0.5\mylenA\usebox\myboxB}%
    \fi}
\makeatother


%% file: NA_v1.bbl
\begin{thebibliography}{88}
\bibitem{White93}
 White, S. D. M.; Navarro, J. F.; Evrard, A. E. \& Frenk, C. S. 
 The baryon content of galaxy clusters: a challenge to cosmological orthodoxy.
 {\it Natur. } {\bf 366}, 429-433 (1993).
 \bibitem{Fukugita98}
Fukugita, M.; Hogan, C. J. \& Peebles, P. J. E. 
The cosmic baryon budget.
{\it Astrophys. J. } {\bf 503}, 518-530 (1998).
\bibitem{Walker12}
Walker, M. G. in {\it Planets Stars and Stellar Systems} Vol. 5 (eds Oswalt, T. D. \& Gilmore, G.) 1039-1089 (2013).
\bibitem{Strigari08}
Strigari, L. E. et al. 
A common mass scale for satellite galaxies of the Milky Way.
{\it Natur. } {\bf 454}, 1096-1097 (2008).
\bibitem{van18}
van Dokkum P. et al.
A galaxy lacking dark matter.
{\it Natur. } {\bf 555}, 629-632 (2018).
\bibitem{van19}
van Dokkum P. et al.
A second galaxy missing dark matter in the NGC 1052 group.
{\it Astrophys. J. } {\bf 874}, L5-L13 (2019).
\bibitem{Oman16}
 Oman, A.-K. et al. 
 Missing dark matter in dwarf galaxies?
{\it Mon. Not. R. Astron. Soc.} {\bf 460}, 3610 (2016)
\bibitem{Planck18}
Planck Collaboration et al.
Planck 2018 results. VI. Cosmological parameters. 
Preprint at http://arxiv.org/abs/1807.06209 (2018).
 \bibitem{Simon07}
 Simon, J. D. \& Geha, M. 
The kinematics of the ultra-faint Milky Way satellites: Solving the missing satellite problem.
{\it Astrophys. J. } {\bf 670}, 313-331 (2007).
\bibitem{Martin07}
Martin, N. F. et al. 
A Keck/DEIMOS spectroscopic survey of faint Galactic satellites: searching for the least massive dwarf galaxies.
{\it Mon. Not. R. Astron. Soc.} {\bf 380}, 281-300 (2007).
\bibitem{Sawala15}
Sawala, T. et al. 
Bent by baryons: the low-mass galaxy-halo relation.
{\it Mon. Not. R. Astron. Soc.} {\bf 448}, 2941-2947 (2015).
\bibitem{Oh15}
 Oh, S.-H. et al. 
 High-resolution mass models of dwarf galaxies from LITTLE THINGS.
{\it Astronomical. J.} {\bf 149}, 180 (2015)
\bibitem{Bournaud07}
Bournaud, F. et al. 
Missing mass in collisional debris from galaxies.
{\it Science.} {\bf 316}, 1166 (2007)
\bibitem{Fritz18}
Fritz, T. K. et al.
Gaia DR2 proper motions of dwarf galaxies within 420 kpc: orbits, Milky Way mass, tidal influences, planar alignments, and group infall. 
Preprint at http://arxiv.org/abs/1805.00908 (2018).
\bibitem{Haynes11}
 Haynes, M. P. et al. 
 The Arecibo Legacy Fast ALFA survey: the $\alpha$.40 HI source catalog, its characteristics and their Impact on the derivation of the HI mass function.
 {\it Astronomical.J.} {\bf 142}, 170 (2011).
\bibitem{Abazajian09}
Abazajian, K. N. et al. 
The seventh data release of the Sloan Digital Sky Survey.
{\it Astrophys. J. Suppl. Ser.} {\bf 182}, 543-558 (2009).
\bibitem{Martin05}
Martin, D. C. et al. 
The Galaxy Evolution Explorer: A space ultraviolet survey mission.
{\it Astrophys. J. } {\bf 619}, L1-L6 (2005).
\bibitem{Lisker2007}
Lisker, T.; Grebel, E. K.; Binggeli, B. \& Katharina, G.
Virgo cluster early-type dwarf galaxies with the Sloan Digital Sky Survey. III. subpopulations: distributions, shapes, origins.
{\it Astrophys. J. } {\bf 660}, 1186-1197 (2007).
\bibitem{Zheng2015}
Zheng, Z. et al.
The structure and stellar content of the outer disks of galaxies: A new view from the Pan-STARRS1 medium deep survey.
{\it Astrophys. J. } {\bf 800}, 120 (2015).
\bibitem{Bell03}
Bell E. F., McIntosh D. H., Katz N. \& Weinberg M. D.
The optical and near-infrared properties of galaxies. I. luminosity and stellar mass functions.
{\it Astrophys. J. Suppl. Ser.} {\bf 149}, 289-312 (2003).
\bibitem{Wang16}
Wang J. et al.
New lessons from the HI size-mass relation of galaxies.
 {\it Mon. Not. R. Astron. Soc.} {\bf 460}, 2143-2151 (2016).
\bibitem{NFW97}
 Navarro J. F.; Frenk C. S. \& White S. D. M. 
 A universal density profile from hierarchical clustering. 
 {\it Astrophys. J.} {\bf 490}, 493-508 (1997)
 \bibitem{Burkert95}
 Burkert, A.
The structure of dark matter halos in Dwarf Galaxies.
{\it Astrophys. J. } {\bf 447}, L25-L28 (1995).
\bibitem{Blanton08}
Blanton, M. R.; Geha, M. \& West, A. A. 
Testing cold dark matter with the low-mass Tully-Fisher relation.
{\it Astrophys. J. } {\bf 682}, 861-873 (2008).
\bibitem{Springob05}
Springob, C. M.; Haynes, M. P.; Giovanelli, R. \& Kent, B. R. 
A digital archive of H I 21 centimeter line spectra of optically targeted galaxies.
{\it Astrophys. J. Suppl. Ser.} {\bf 160}, 149-162 (2005).
 \bibitem{Jing19}
 Jing. Y. J. et al.
 The dark matter deficit galaxies in hydrodynamical simulations.
Preprint at https://arxiv.org/abs/1811.09070 (2018).
\bibitem{Ogiya18}
Haslbauer, M. et al.
Galaxies lacking dark matter in the Illustris simulation.
{\it Astron. Astrophys.} {\bf 626}, A47 (2019).
\bibitem{Hunter12}
 Hunter, D. A. et al. 
 Little things.
{\it Astronomical. J.} {\bf 144}, 134 (2012)
\bibitem{Vogelsberger2014}
Vogelsberger, M. et al.
Introducing the Illustris Project: simulating the coevolution of dark and visible matter in the Universe
{\it Mon. Not. R. Astron. Soc.} {\bf 444}, 1518-1547 (2014).
\bibitem{Diaz-Garcia2016}
Diaz-Garcia, S.; Salo, H.; Laurikainen, E. \& Herrera-Endoqui, M.
{\it Astronomy \& Astrophysics} {\bf 587}, A160-A220 (2016)
\end{thebibliography}

\begin{thebibliography}{30}
\setcounter{enumiv}{28}

\bibitem[31]{Giovanelli05a}
Giovanelli, R. et al.
The Arecibo Legacy Fast ALFA Survey. I. science goals, survey design, and strategy.
{\it Astronomical.J.} {\bf 130}, 2598 (2005).
\bibitem[32]{Giovanelli07}
Giovanelli, R. 
ALFALFA: An exploration of the z=0 HI universe.
{\it Il. Nuovo. Cimento. B} {\bf 122}, 1097-1108 (2007).
\bibitem[33]{Haynes07}
Haynes, M. P. 
ALFALFA: The search for (almost) dark galaxies and their space distribution.
{\it Il. Nuovo. Cimento. B} {\bf 122}, 1109-1114 (2007).
\bibitem[34]{Giovanelli05b}
Giovanelli, R. et al.
The Arecibo Legacy Fast ALFA Survey. II. results of precursor observations.
{\it Astronomical.J.} {\bf 130}, 2613 (2005).
 \bibitem[35]{Catinella12}
 Catinella, B. et al.
 The GALEX Arecibo SDSS Survey. VI. Second data release and updated gas fraction scaling relations.
{\it A. \&A.} {\bf 544}, A65 (2012). 
\bibitem[36]{He2013}
 He, Y. Q. et al. 
 Photometric properties and luminosity function of nearby massive early-type galaxies.
{\it Astrophys. J. } {\bf 773}, 37 (2013).
\bibitem[37]{Liu2008}
Liu, F. S.; Xia, X. Y.; Mao, S.; Wu, H. \& Deng, Z. G.
Photometric properties and scaling relations of early-type brightest cluster galaxies.
{\it Mon. Not. R. Astron. Soc.} {\bf 385}, 23-29 (2008).
\bibitem[38]{Du15}
Du, Wei et al.
Low surface brightness galaxies selected from the 40\% sky area of the ALFALFA H I survey. I. sample and statistical properties.
{\it Astronomical.J.} {\bf 149}, 199 (2015).
\bibitem[39]{Alam15}
Alam, S. et al. 
The eleventh and twelfth data releases of the Sloan Digital Sky Survey: final data from SDSS-III.
{\it Astrophys. J. Suppl. Ser.} {\bf 219}, 12 (2015).
\bibitem[40]{Bertin1996}
Bertin, E. \& Arnouts. 
SExtractor: Software for source extraction.
{\it A.\&A. S.} {\bf 117}, 393-404 (1996).
 \bibitem[41]{Blanton2001}
 Blanton, M. R. et al.
 The luminosity function of galaxies in SDSS commissioning data.
 {\it Astronomical. J.} {\bf 121}, 2358-2380 (2001).
 \bibitem[42]{Yasuda2001}
 Yasuda, N. et al.
 Galaxy number counts from the Sloan Digital Sky Survey commissioning data.
 {\it Astronomical. J.} {\bf 122}, 1104-1124 (2001).
\bibitem[43]{Roberts62}
Roberts, M. S.
The neutral hydrogen content of late-type spiral galaxies.
 {\it Astronomical. J.} {\bf 67}, 437-446 (1962).
 \bibitem[44]{Masters05}
Masters, K. L.
Galaxy flows in and around the Local Supercluster.
 {\it Ph.D dissertation} {\bf 0058}, 0606 (2005) AAT 3192074.
 \bibitem[45]{Kroupa03}
 Kroupa, P.
The initial mass function of stars: Evidence for uniformity in variable systems.
{\it Science.} {\bf 295}, 82-91 (2002)
\bibitem[46]{Giovanelli97}
Giovanelli, R. et al. 
The I band Tully-Fisher relation for cluster galaxies: data presentation.
{\it Astronomical. J.} {\bf 113}, 22-52 (1997)
\bibitem[47]{Tully09}
Tully, B. R. et al. 
The extragalactic distance database.
{\it Astronomical. J.} {\bf 138}, 323-331 (2009)
\bibitem[48]{Starkenburg2019}
Starkenburg S. et al.
On the origin of star-gas counterrotation in low-mass galaxies.
{\it Astrophys. J.} {\bf 878}, 143-155 (2019).
\bibitem[49]{Dutton14}
Dutton A. A. \& Macci$\small\grave{o}$ A. V.
Cold dark matter haloes in the Planck era: evolution of structural parameters for Einasto and NFW profiles.
 {\it Mon. Not. R. Astron. Soc.} {\bf 441}, 3359-3374 (2014).
 \bibitem[50]{Salucci07}
Salucci P.  et al.
The universal rotation curve of spiral galaxies – II. The dark matter distribution out to the virial radius.
 {\it Mon. Not. R. Astron. Soc.} {\bf 378}, 41-44 (2007).
  \bibitem[51]{Huchra12}
 Huchra, J. P. et al.
The 2MASS Redshift Survey -- description and data release.
 {\it Astrophys. J. Suppl. Ser.} {\bf 199}, 26 (2012)
  \bibitem[52]{Saulder16}
 Saulder, C.; van Kampen, E.; Chilingarian, I.; Mieske, S. \& Zeilinger, W. W. 
The matter distribution in the local Universe as derived from galaxy groups in SDSS DR12 and 2MRS.
{\it A.\&A.} {\bf 596}, A14 (2016).
 \bibitem[53]{Guo11}
Guo, Qi et al.
From dwarf spheroidals to cD galaxies: simulating the galaxy population in a $\rm \Lambda CDM$  cosmology.
 {\it Mon. Not. R. Astron. Soc.} {\bf 413}, 101-131 (2011).
  \bibitem[54]{Guo13}
Guo, Qi et al.
Galaxy formation in WMAP1 and WMAP7 cosmologies.
 {\it Mon. Not. R. Astron. Soc.} {\bf 428}, 1351-1365 (2013).
 \bibitem[55]{Rines03}
 Rines, K.; Geller, M.; Kurtz, M. \& Diaferio, A.
 CAIRNS: The Cluster and Infall Region Nearby Survey. I. Redshifts and Mass Profiles.
 {\it Astronomical. J.} {\bf 126}, 2152-2170 (2003).
\end{thebibliography}
